\pdfoutput=1
\documentclass[twocolumn,epjc3]{svjour3}

\RequirePackage[T1]{}

\usepackage{amsmath}
\usepackage{enumitem}
\usepackage{hyperref}

\smartqed

\RequirePackage{graphicx}
\RequirePackage{mathptmx}
\RequirePackage{flushend}
\usepackage{multirow}
\usepackage{amssymb}
\usepackage{amsmath}
\usepackage{romannum}
\RequirePackage[numbers,sort&compress]{natbib}

\journalname{}

\begin{document}
			
\title{ The Standard siren tests of viable $f(R)$ cosmologies  }

\author{
		Yi Zhang\thanksref{e1,addr1}
		\and
		Xuanjun Niu\thanksref{addr1}
		\and
		Xianfu Su\thanksref{addr2}
		\and
		Dong-Ze He\thanksref{addr1}}

\thankstext{e1}{e-mail: zhangyia@cqupt.edu.cn}
\institute{School of Electronic Science and Engineering,Chongqing University
				of Posts and Telecommunications, Chongqing 400065, China\label{addr1}
				\and
				Zhijin No.9 Senior High School,Guizhou 552100, China\label{addr2}}
			
\date{Received: date / Accepted: date}

\maketitle
	
\begin{abstract}
	We constrain the Hu-Sawicki and Starobinsky $f(R)$ gravity models utilizing current electromagnetic   (PP+CC, Planck and DESI2) datasets and simulate standard siren  catalogs based on the resulting best-fit parameters. We demonstrate that the simulated SS data provide complementary sensitivity to the modified gravitational wave propagation friction term, thereby enhancing the discriminating power between $f(R)$ gravity and the $\Lambda$CDM model. However, we note that standard sirens do not offer a viable resolution to the Hubble tension in this analysis, as the inferred constraints are predominantly driven by the fiducial cosmologies adopted in the simulations. Regarding the specific models, we find that for the Hu-Sawicki scenario, several data combinations favor $F_{RR0}<0$, implying potential theoretical instabilities. And, for the Starobinsky model, while EM-only constraints are nearly symmetric between the two parameter branches ($b<0$ and $b>0$), the inclusion of SS constraints introduces mild asymmetries, revealing the sensitivity of SS observables to the curvature dependence of the theory. Future truly independent standard siren observations would
	be crucial for a definitive assessment of $f(R)$ gravity as an
	alternative to $\Lambda$CDM.
\end{abstract}


\section{Introduction}
Recent baryon acoustic oscillation (BAO) measurements from the Dark Energy
Spectroscopic Instrument (DESI)~\cite{DESI:2025zgx,DESI:2025fii}, especially
when combined with cosmic microwave background (CMB) and Type~Ia supernova
(SNe~Ia) observations, suggest a mild preference for departures from a pure
cosmological constant and therefore motivate tests of a dynamical dark-energy
sector. At the same time, despite extensive theoretical efforts, many
cosmological models remain challenged by the Hubble tension in electromagnetic
(EM) data, commonly defined as a $>5\sigma$ discrepancy between local
distance ladder and CMB inferred determinations of the Hubble constant
$H_0$~\cite{DiValentino:2020zio,Abdalla:2022yfr,Planck:2018vyg,Riess:2021jrx,Breuval:2024lsv,Murakami:2023xuy}.
Such tensions may arise from unaccounted systematics or, more intriguingly,
from physics beyond the standard $\Lambda$CDM framework. These considerations
provide strong motivation to test modified gravity scenarios as alternatives to
$\Lambda$CDM model.

Among modified gravity theories, $f(R)$ gravity offers a natural and minimal
extension of general relativity (GR) that can account for late-time cosmic
acceleration without invoking a fundamental cosmological constant, while still
admitting a continuous $\Lambda$CDM limit. Phenomenologically, this can be
interpreted as an effective dark energy equation of state $w_{\rm eff}(z)$ that deviates
from $-1$ and may evolve with redshift $z$. 
However, when constrained solely by EM probes of the background expansion, most
viable $f(R)$ models are notoriously difficult to distinguish from
$\Lambda$CDM~\cite{Nunes:2016drj,Farrugia:2021zwx,DAgostino:2020dhv,Basilakos:2013nfa,Matos:2021qne,Leizerovich:2021ksf,Bessa:2021lnr,Jana:2018djs,Amendola:2006we,Capozziello:2007eu,Fu:2010zza,Hu:2016zrh,Amendola:2006kh,Odintsov:2020qzd,Kumar:2025mzo,He:2026yhe}.
This motivates the use of complementary observables with enhanced sensitivity
to modified gravity effects. A key question is therefore whether
$f(R)$ gravity can provide a viable alternative to $\Lambda$CDM when confronted
with forthcoming standard siren (SS) tests.

Gravitational waves (GWs), a cornerstone prediction of GR, provide such an
independent probe. Since the first detection of the binary black hole merger
GW150914, hundreds of compact-binary coalescences have been observed
\cite{LIGOScientific:2016aoc,LIGOScientific:2017vwq,LIGOScientific:2017zic,LIGOScientific:2018mvr,LIGOScientific:2020ibl,LIGOScientific:2025slb}.
In particular, standard sirens  yield a direct measurement of the
luminosity distance $D_L^{\rm SS}$ without relying on the cosmic distance
ladder. In modified gravity, the SS luminosity distance can differ from its EM
counterpart due to an effective friction term in GW propagation, providing a
powerful discriminator between modified gravity and $\Lambda$CDM
\cite{Belgacem:2017ihm,Belgacem:2019pkk}. Although the current number of SS
events is still limited, future detectors such as the Einstein Telescope (ET)
are expected to observe large samples of binary neutron star mergers (including
lensed events), enabling high-precision tests of GR and $\Lambda$CDM
\cite{Holz:2005df,Schutz:1986gp}. Following the simulation frameworks of
Refs.~\cite{gwbook,Maggiore2007,Zhao:2010sz,Cai:2016sby,Su:2024avk,Zhang:2021kqn},
we will generate mock ET standard siren catalogs to assess whether SS observations
can break the degeneracy between $f(R)$ gravity and $\Lambda$CDM here.

In this work we focus on two representative and widely studied $f(R)$ models:
the Hu-Sawicki and Starobinsky models~\cite{Hu:2007nk,Starobinsky:2007hu},
which satisfy standard viability conditions and recover $\Lambda$CDM in an
appropriate limit. Motivated by the bimodal posteriors discussed in
Ref.~\cite{Basilakos:2013nfa}, we further split the Starobinsky model into two
branches with $b>0$ and $b<0$. Since EM data (SNe~Ia, BAO, and CMB) alone do not
provide decisive evidence for dynamical dark energy, we use the EM best fitted
results as fiducial inputs for our SS simulations in order to isolate and
quantify the additional impact of standard sirens.

This paper is organized as follows. In Sec.~\ref{fR} we introduce the
Hu-Sawicki and Starobinsky models. In Sec.~\ref{data} we describe the EM and
standard siren datasets used in the analysis. The cosmological constraints and
discussion are presented in Sec.~\ref{Discussion}. Finally, we summarize our
conclusions in Sec.~\ref{con}.


\section{The $f(R)$ models}\label{fR}

In the metric formalism, the Einstein-Hilbert action is generalized by
replacing the Ricci scalar $R$ with a generic function $f(R)$. A general
$f(R)$ theory minimally coupled to matter is described by
\begin{eqnarray}
	\label{action}
	S=\int d^{4}x\,\sqrt{-g}\left[\frac{f(R)}{16\pi G}+\mathcal{L}_{m}\right],
\end{eqnarray}
where $\mathcal{L}_{m}$ denotes the Lagrangian of the non-gravitational matter
sector, including dark matter.\footnote{We neglect radiation, which is
	subdominant at late times.} The function $f(R)$ encodes deviations from general
relativity (GR) and may account for the late-time cosmic acceleration. It is
convenient to decompose
\begin{eqnarray}
	f(R)=R+F(R),
\end{eqnarray}
so that $F(R)$ measures the departure from GR. Varying the action with respect
to the metric yields the modified Einstein equations. For a spatially flat
Friedmann-Robertson-Walker (FRW) background, the Ricci scalar is
\begin{eqnarray}
	R=12H^{2}+6\dot{H},
\end{eqnarray}
and the modified Friedmann equations can be written as
\begin{eqnarray}
	\label{friedmann1}
&&	3 f_{R} H^{2}-\frac{R f_{R}-f(R)}{2}+3H f_{RR}\dot{R} = 8\pi G\,\rho_{m},\\
	\label{friedmann2}
&&	-2 f_{R}\dot{H} = 8\pi G\,\rho_{m}+\ddot{f}_{R}-H\dot{f}_{R},
\end{eqnarray}
where overdots denote derivatives with respect to cosmic time $t$,
$\rho_{m}$ is the matter energy density, and $f_{R}\equiv df/dR$,
$f_{RR}\equiv d^{2}f/dR^{2}$.

The coupled system in Eqs.~(\ref{friedmann1}) and (\ref{friedmann2}) typically does
not admit closed-form solutions. In practice, viable models such as the
Hu-Sawicki and Starobinsky forms are constructed to smoothly approach
$\Lambda$CDM in an appropriate limit. Their functional form can be cast as
\begin{eqnarray}
	\label{fRy}
	f(R)=R-2\Lambda\,y(R,b),
\end{eqnarray}
where $\Lambda$ is an effective cosmological constant and $y(R,b)$ parametrizes
the deviations from $\Lambda$CDM via a dimensionless parameter $b$. For $b>0$,
one has $\lim_{b\to 0}f(R)=R-2\Lambda$ and $\lim_{b\to\infty}f(R)=R$. In
particular, as $F_{R}\to 0$ and $F_{RR}\to 0$ the theory continuously reduces
to $\Lambda$CDM. Although originally proposed as cosmological constant free
models, both the Hu-Sawicki and Starobinsky constructions effectively contain
a $\Lambda$ term in the relevant limit~\cite{Starobinsky:2007hu,Hu:2007nk}.

Defining the normalized Hubble rate $E(a)\equiv H/H_{0}$, an effective
dark-energy equation of state can be expressed as
\begin{eqnarray}
	w_{\rm eff}(a)=\frac{-1-\frac{2a}{3}\frac{d\ln E}{da}}{1-\Omega_{m}(a)},
\end{eqnarray}
where $\Omega_{m}(a)=\Omega_{m0}a^{-3}/E^{2}(a)$.
In the $\Lambda$CDM limit, $w_{\rm eff}=-1$.

Following Basilakos et al.~\cite{Basilakos:2013nfa}, we adopt an efficient
approximation scheme based on expanding the Hubble rate around the $\Lambda$CDM
solution, which avoids expensive numerical integrations and has been widely
used~\cite{Nunes:2016drj,Farrugia:2021zwx,DAgostino:2019hvh,DAgostino:2020dhv}.
Rewriting Eq.~(\ref{friedmann1}) in terms of the e-folding number $N=\ln a$,
we obtain
\begin{eqnarray}
	\label{Nexpansion}
	-f_{R}H^{2}(N)+\Omega_{m0}e^{-3N}+\frac{1}{6}\bigl(R f_{R}-f\bigr)
	=f_{RR}H^{2}(N)\,R'(N),
\end{eqnarray}
where the prime ``$\prime$" denotes derivative with respect to $N$. To quantify deviations
from $\Lambda$CDM we expand around $b=0$; keeping the first two nonzero terms
reproduces the numerical solution to better than $0.001\%$ for
$b\in[0.001,0.5]$. Specifically,
\begin{eqnarray}
	\label{expansion1}
	&&H^{2}(N)=H_{\Lambda}^{2}(N)+\sum_{i=1}^{M} b^{i}\,\delta H_{i}^{2}(N),
	\\
	&&H_{\Lambda}^{2}(N)=\Omega_{m0}e^{-3N}+\bigl(1-\Omega_{m0}\bigr),
\end{eqnarray}
where $M$ denotes the truncation order.

The dynamical properties of $f(R)$ models can be conveniently characterized by
two dimensionless variables~\cite{Amendola:2006we,Tsujikawa:2007xu},
\begin{eqnarray}
	\label{r}
	&&r\equiv -\frac{R f_{R}}{f(R)}=-\frac{R\bigl(1+F_{R}\bigr)}{f(R)},\\
	\label{m}
	&& m\equiv \frac{R f_{RR}}{f_{R}}
	=\frac{R\,F_{RR}}{1+F_{R}}.
\end{eqnarray}
Here, $f_{R}=1+F_{R}$ and $f_{RR}=F_{RR}$. The quantity $f_{R}$ controls the
effective gravitational coupling,
$G_{\rm eff}\propto f_{R}^{-1}$, whereas $f_{RR}$ governs the dynamics of the
additional scalar degree of freedom (the scalaron) and thus sets how rapidly
deviations from GR/$\Lambda$CDM can develop. The variable $r$ measures the
proximity to the $\Lambda$CDM limit, while the variable $m$ quantifies the strength of the
deviation; in many dynamical analyses $m<0$ signals an instability
\cite{Amendola:2006we}. Viable cosmologies correspond to trajectories in the
$(r,m)$ plane that pass close to the GR limit $(r,m)\simeq(-1,0^{+})$ during the
radiation and matter eras and then approach a stable de~Sitter fixed point near
$r\simeq -2$ with $0<m<1$. Within this framework, departures from
$\Lambda$CDM, for example $w_{\rm eff}(z)\neq -1$ and its redshift
evolution, can be interpreted as resulting from the evolution of the
$(r,m)$ variables (equivalently, from the curvature dependence encoded in
$m(R)$).

\subsection{The Hu-Sawicki model}

Hu and Sawicki~\cite{Hu:2007nk} proposed a viable $f(R)$ model of the form
\begin{eqnarray}
	\label{hu}
	F(R)=-\tilde{m}^2\frac{c_1\left(R/\tilde{m}^2\right)^n}{1+c_2\left(R/\tilde{m}^2\right)^n},
\end{eqnarray}
where $\tilde{m}^2=\Omega_{m0}H_0^2$,  $c_1$, $c_2$, and $n$ are free
parameters. Equivalence-principle tests typically require $n\gtrsim 0.9$
\cite{Capozziello:2007eu}; in this work we restrict to $n=1$. For $n=1$, the
model can be rewritten as
\begin{eqnarray}
	\label{hueq}
	F(R)=-\frac{2\Lambda}{1+b\Lambda/R},
\end{eqnarray}
where $b\equiv 2/c_1$ and $\Lambda=\tilde{m}^2 c_1/(2c_2)$. The Hu-Sawicki
model can be made arbitrarily close to $\Lambda$CDM by taking $b\to 0$. In this
case, Eq.~(\ref{fRy}) yields
\begin{eqnarray}
	\label{y}
	y_{\rm HS}(R,b)=1-\frac{1}{1+R/(b\Lambda)}.
\end{eqnarray}

For later use, the first two derivatives of Eq.~(\ref{hueq}) with respect to
$R$ are list as
\begin{eqnarray}
	F_R=-\frac{2}{b\left(1+R/(b\Lambda)\right)^2},
	\\
	F_{RR}=\frac{4}{b^2\Lambda\left(1+R/(b\Lambda)\right)^3},
\end{eqnarray}
where $F_R$ is an old function, while $F_{RR}$ is an even one.

Expanding the Hubble rate according to Eq.~(\ref{expansion1}), we obtain
\begin{eqnarray}
	\label{expansionHS1}
	H_{\rm HS}^2(N)=H_{\Lambda}^2(N)+b\,\delta H_1^2(N)+b^2\,\delta H_2^2(N)+\cdots,
\end{eqnarray}
where $\delta H_1^2(N)$ and $\delta H_2^2(N)$ are given in
Ref.~\cite{Basilakos:2013nfa}. The leading departure from $\Lambda$CDM appears
at ${\cal O}(b)$, i.e.\ this model is $\Lambda$-like in the small-$b$ regime.

\subsection{The Starobinsky model}

The Starobinsky model~\cite{Starobinsky:2007hu} is defined by
\begin{eqnarray}
	F(R)=-c_1\tilde{m}^2\left[1-\left(1+R^2/\tilde{m}^4\right)^{-n}\right],
\end{eqnarray}
where $\tilde{m}^2=\Omega_{m0}H_0^2$, and $c_1$ and $n$ are free parameters. It
has been argued that $n$ should be an integer~\cite{Fu:2010zza}; therefore we
set $n=1$ following Ref.~\cite{Basilakos:2013nfa}. For $n=1$, the model can be
written as
\begin{eqnarray}
	\label{Star}
	F(R)=-2\Lambda\,
	\frac{\left(R/(b\Lambda)\right)^2}{1+\left(R/(b\Lambda)\right)^2},
\end{eqnarray}
where $\Lambda=c_1\tilde{m}^2/2$ and $b\equiv 2/c_1$. The model approaches
$\Lambda$CDM in the limit $b\to 0$. The corresponding deviation function is
\begin{eqnarray}
	\label{stargr}
	y_{\rm Star}(R,b)=1-\frac{1}{1+\left(R/(b\Lambda)\right)^2}.
\end{eqnarray}

The first two derivatives of Eq.~(\ref{Star}) are
\begin{eqnarray}
	F_R=-\frac{4R}{b^2\Lambda\left[1+\left(R/(b\Lambda)\right)^2\right]^2},
	\\
	F_{RR}=\frac{4}{b^2\Lambda}\,
	\frac{3\left(R/(b\Lambda)\right)^2-1}{\left[1+\left(R/(b\Lambda)\right)^2\right]^3},
\end{eqnarray}
which are two even functions.
Stability requires $\tilde{m}^2>0$ for viable $f(R)$ models~\cite{Yang:2011cp}.
Expanding the Friedmann equation yields
\begin{eqnarray}
	\label{expansionstar1}
	H_{\rm Star}^2(N)=H_{\Lambda}^2(N)+b^2\,\delta H_2^2(N)+b^4\,\delta H_4^2(N)+\cdots,
\end{eqnarray}
where the leading correction appears at ${\cal O}(b^2)$. The explicit forms of
$\delta H_2^2(N)$ and $\delta H_4^2(N)$ are given in
Ref.~\cite{Basilakos:2013nfa} as well.

For the Starobinsky case, EM constraints on $b$  may
exhibit a bimodal structure~\cite{Kumar:2025mzo}. To capture this feature, we
impose the prior $b\neq 0$, which partitions the parameter space into two
branches, $b<0$ and $b>0$, denoted as Starobinsky\Romannum{1} and
Starobinsky\Romannum{2} (Star\Romannum{1} and Star\Romannum{2} for short). 

\section{Data and methodology}\label{data}

We constrain the parameters of the $f(R)$ models using a Markov Chain Monte
Carlo (MCMC) analysis together with a maximum-likelihood estimation, implemented
with \texttt{CosmoMC}~\cite{Lewis:2002ah}. In this section, we shall describe the
electromagnetic  datasets and the simulated standard siren  data used
in our analysis.

\subsection{Electromagnetic datasets}\label{sec:emdata}

We consider EM probes of the cosmic distance scale, including measurements of
the Hubble parameter from cosmic chronometers (CC)~\cite{Moresco:2020fbm,Favale:2023lnp},
Type~Ia supernovae from the PantheonPlus (PP) compilation~\cite{Scolnic:2021amr,Brout:2022vxf},
CMB distance priors from Planck~\cite{Planck:2018vyg}, and BAO measurements from
the  DESI Data Release~2 (hereafter DESI2)~\cite{DESI:2025zgx,DESI:2025fii}.
The PP and CC data mainly probe the late-time Universe, whereas CMB and BAO
measurements are also sensitive to earlier epochs.

\subsubsection{PantheonPlus + cosmic chronometers (PP+CC) }\label{sec:ppcc}

The PantheonPlus compilation provides 1701 distance modulus measurements from
18 independent surveys~\cite{Scolnic:2021amr,Brout:2022vxf} over the redshift
range $0.00122<z<2.26137$. For PantheonPlus, we compute the likelihood using
the full covariance matrix $\mathbf{C}_{\rm stat+sys}$, which includes both
statistical and systematic uncertainties.

The CC data estimate $H(z)$ from the differential ages of passively evolving
galaxies. We adopt the compilation of 32 CC measurements listed in Table~1 of
Ref.~\cite{Favale:2023lnp}, including determinations based on galaxy age dating
and BAO-related techniques~\cite{Zhang:2012mp,Jimenez:2003iv,Moresco:2012by,Moresco:2015cya,Simon:2004tf,Chuang:2011fy,Moresco:2016mzx,Stern:2009ep,Borghi:2021rft}.

Since the PP and CC datasets are mutually consistent~\cite{Sasi:2024vwl}, we
combine them into a single dataset denoted PP+CC. The covariance matrices used
in the likelihood follow Refs.~\cite{Moresco:2020fbm,Gomez-Valent:2018hwc,Chuang:2011fy}.
The combined PP+CC dataset contains 1733 data points.

\subsubsection{Planck distance prior }\label{sec:planck}

For CMB constraints, we use the distance priors derived from the Planck 2018
data release~\cite{Planck:2018vyg}. The distance prior approach
\cite{Hu:1995en,Efstathiou:1998xx,Wang:2006ts,Wang:2007mza,Chen:2018dbv,Zhai:2019nad,Zhai:2018vmm}
provides a compressed representation of the CMB information while retaining the
key cosmological constraints relevant for background evolution. The priors are
typically expressed in terms of the shift parameter $R$ and the acoustic scale
$\ell_a$~\cite{Hu:1995en}. We adopt the data vector and covariance matrix from
Ref.~\cite{Zhai:2018vmm}.

\subsubsection{DESI2 BAO }\label{sec:desi2}

Baryon acoustic oscillations  provide a standard ruler for tracing the
expansion history through features imprinted in the matter power spectrum. We
use the DESI  BAO release 2 (DESI2)~\cite{DESI:2025zgx,DESI:2025fii}, which includes measurements from four classes of tracers: the bright
galaxy sample (BGS)~\cite{Hahn:2022dnf}, luminous red galaxies (LRG)~\cite{DESI:2022gle},
emission line galaxies (ELG)~\cite{Raichoor:2022jab}, and quasars (QSO)~\cite{Chaussidon:2022pqg}.

DESI2 reports the transverse comoving distance $D_M$, the Hubble distance
$D_H$, and the volume-averaged distance $D_V$, normalized by the sound horizon
at the drag epoch $r_d$. We determine $r_d$ using BBN information.\footnote{We
	adopt $\Omega_b h^2=0.022$~\cite{Cooke:2017cwo}, where $\Omega_b$ is the baryon
	density parameter and $h\equiv H_0/(100~{\rm km\,s^{-1}\,Mpc^{-1}})$.}
The measurements span seven redshift bins derived from over six million objects
(see Table~1 of Refs.~\cite{DESI:2025zgx,DESI:2025fii}).

For LRG1, LRG2, LRG3+ELG1, ELG2, and Ly$\alpha$ QSO, DESI2 provides correlated
pairs of $(D_M/r_d,\; D_H/r_d)$. Following Ref.~\cite{Li:2024bwr}, we construct
the corresponding data vector and covariance matrix, using the correlation
coefficients reported in Table~1 of Ref.~\cite{DESI:2025zgx}.



\subsection{Standard sirens}\label{standard}

Gravitational waves  emitted by compact binary coalescences can serve as the
standard sirens to probe the cosmic expansion history
\cite{Holz:2005df,Schutz:1986gp}. In particular, the luminosity distance
$D_L^{\rm SS}$ can be inferred directly from the GW amplitude, without
calibration from the cosmic distance ladder,
\begin{eqnarray}
	h_A=\frac{4}{D_L^{\rm SS}}
	\left(\frac{G M_c}{c^2}\right)^{5/3}
	\left(\frac{\pi f_{\rm SS}}{c}\right)^{2/3},
\end{eqnarray}
where $h_A$ is the GW amplitude, $M_c$ is the chirp mass, and $f_{\rm SS}$ is
the GW frequency.

In modified gravity theories, GW propagation generally follows
\cite{Cai:2016sby,Zhang:2021kqn}
\begin{eqnarray}
	\bar h_A''+2\mathcal{H}\bigl[1+\delta(\eta)\bigr]\bar h_A' + k^2 \bar h_A = 0,
\end{eqnarray}
where the $\bar h_A$ denotes the Fourier mode of the GW amplitude, prime ``$\prime$" denotes
derivatives with respect to conformal time $\eta$, $\mathcal{H}\equiv a'/a$ is
the conformal Hubble parameter, and $\delta$ is an extra friction term that
vanishes in GR. Defining an effective scale factor $\tilde a$ via
\begin{eqnarray}
	\frac{\tilde a'}{\tilde a}=\mathcal{H}\bigl[1+\delta(z)\bigr],
\end{eqnarray}
and introducing $\chi_A\equiv \tilde a\,\bar h_A$, the propagation equation can
be recast as~\cite{Belgacem:2017ihm}
\begin{eqnarray}
	\chi_A''+\left(k^2-\frac{\tilde a''}{\tilde a}\right)\chi_A=0.
\end{eqnarray}
As a consequence, the SS and EM luminosity distances are related by
\begin{eqnarray}
	D_L^{\rm SS}(z)=
	\exp\!\left[\int_0^z \frac{\delta(z')}{1+z'}\,dz'\right]\,D_L^{\rm EM}(z).
\end{eqnarray}

In $f(R)$ gravity, the friction term is given by~\cite{Hwang:2001qk}
\begin{eqnarray}
	\delta=-\frac{\dot R\,F_{RR}}{2H\bigl(1+F_R\bigr)},
\end{eqnarray}
which leads to the compact relation
\begin{eqnarray}
	\label{betaT}
	\beta_T(z)\equiv \frac{D_L^{\rm SS}(z)}{D_L^{\rm EM}(z)}
	=\sqrt{\frac{1+F_{R0}}{1+F_R(z)}}.
\end{eqnarray}
By construction, $\beta_T(0)=1$, and the departure of GW distances from EM
distances is fully determined by the redshift evolution of $F_R$.


\subsection{The method}\label{method}
We first constrain the cosmological models using the EM datasets PP+CC, Planck,
and DESI2. The Markov chain Monte Carlo (MCMC) analysis is performed with
\texttt{CosmoMC}~\cite{Lewis:2002ah}. For each EM dataset, the best-fit
parameter values (corresponding to the minimum $\chi^2$) are adopted as the
fiducial cosmology to generate mock standard siren (SS) catalogs. In our
notation, the subscript ``PP+CC/Planck/DESI2'' indicates that the SS fiducial
parameters are taken from the best-fit constraints obtained with the
corresponding EM dataset. Following
Refs.~\cite{gwbook,Zhao:2010sz,Cai:2016sby}, we simulate 1000 SS events
detectable by the Einstein Telescope (ET) over a 10-year observation period,
including 500 binary neutron star (BNS) and 500 black hole-neutron star (BHNS)
systems. The weak-lensing uncertainty is modeled as $\sigma_{\rm lens}=0.05z$.

For most   simulations, we directly adopt the
corresponding best-fit values as fiducial parameters.
However, for the Hu-Sawicki model, the PP+CC best-fit values
$(\Omega_{m0},H_0,b)=(0.252,\,78.65,\,0.62)$ deviate substantially from the
$\Lambda$CDM results. Directly using this parameter set as SS fiducials can
lead to unphysical and/or unstable constraints. To avoid this issue, we instead
use the $\Lambda$CDM-based SS$_{\rm PP+CC}$ mock catalog to constrain the
Hu-Sawicki model, adopting the fiducial parameters
\begin{eqnarray}
	\Omega_{m0}=0.348,\,\, H_0=72.99,\,\, b=0,
\end{eqnarray}
which lie within the $2\sigma$ region of the PP+CC constraints for the
Hu-Sawicki model. In particular, we denote this dataset as
HS:SS$_{\Lambda{\rm CDM:PP+CC}}$.

To quantify the Hubble tension, we use \cite{Raveri:2018wln}
\begin{eqnarray}
	T_{1}(\theta)=\frac{|\theta(D_{1})-\theta(D_{2})|}{\sqrt{\sigma_\theta^{2}(D_1)+\sigma_\theta^{2}(D_2)}},
\end{eqnarray}
where $\theta$ is the best fitted values of $H_0$ from different data sets; the first data set $D_1$ represents the constraining results of cosmological fitting; the second data set $D_2$ is the chosen baseline measurement, which is $H_{0} = 73.17 \pm 0.86 ~\mathrm{km/s/Mpc}$ from the latest SH0ES Team \cite{Riess:2021jrx,Breuval:2024lsv,Murakami:2023xuy}; and   $\sigma_\theta(D_1)$ and $\sigma_\theta(D_2)$ represent the  errors from $D_1$ and $D_2$ data sets respectively. 

Finally, we assess model performance using the AIC and BIC criteria with the $\Lambda\mathrm{CDM}$   model as reference. 
And, for Gaussian errors, the difference between two models could be written as  $\Delta AIC = \Delta \chi^2 + 2\Delta N$. 
Similar to the AIC, the difference denoted by BIC has the form 
$\Delta  BIC = \Delta  \chi^2+ \Delta  N \ln{m}$, where $\Delta N$ is the number of additional parameters and $m$ is the number of data points. The $\Delta AIC=5$ ($\Delta BIC\geq2$ ) and $\Delta AIC=10$ ($\Delta BIC\geq6$) are considered to be the positive and strong evidence against the weaker model respectively.
\begin{figure*}[!htb]
	\centering
	\includegraphics[width=8.65cm]{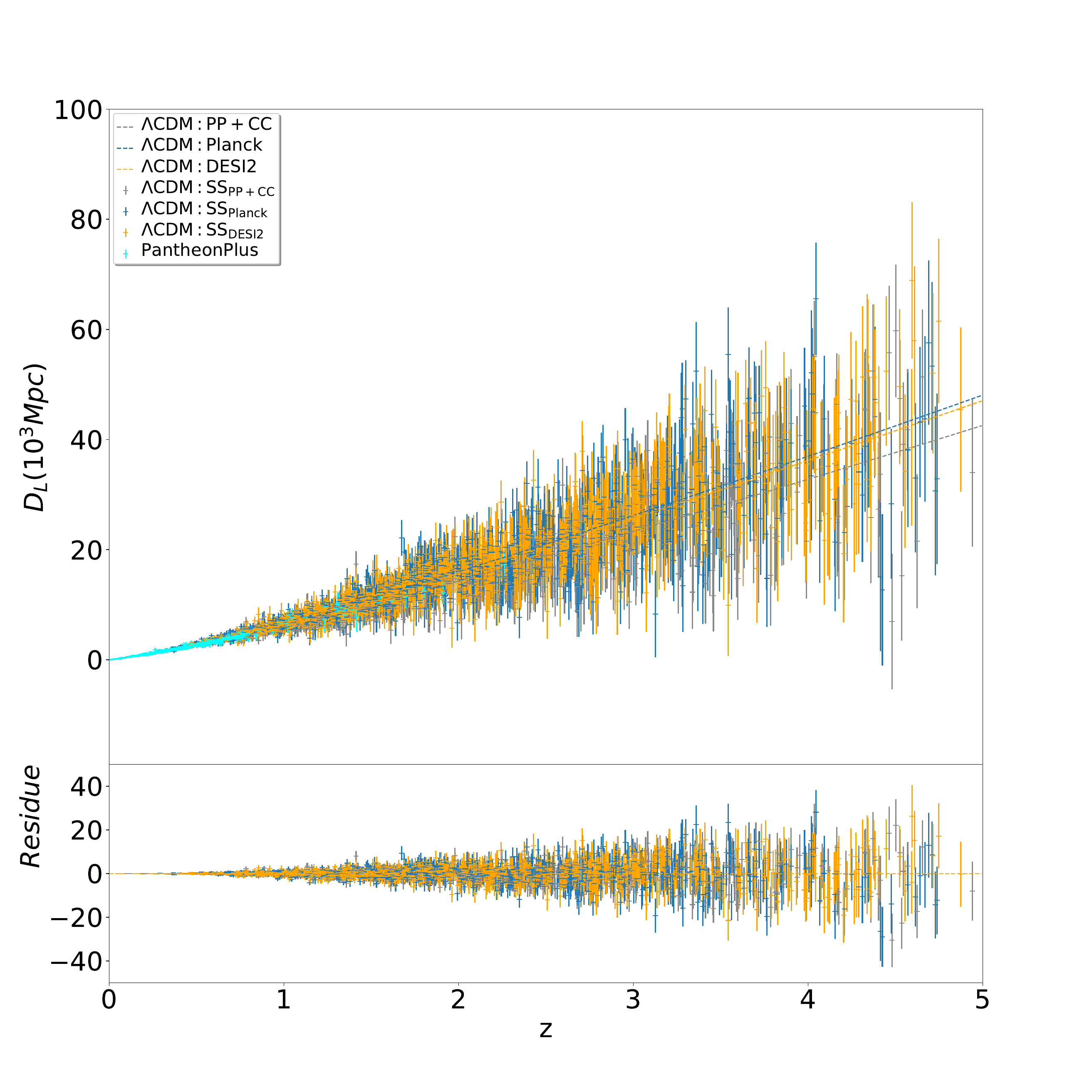}
		\includegraphics[width=8.65cm]{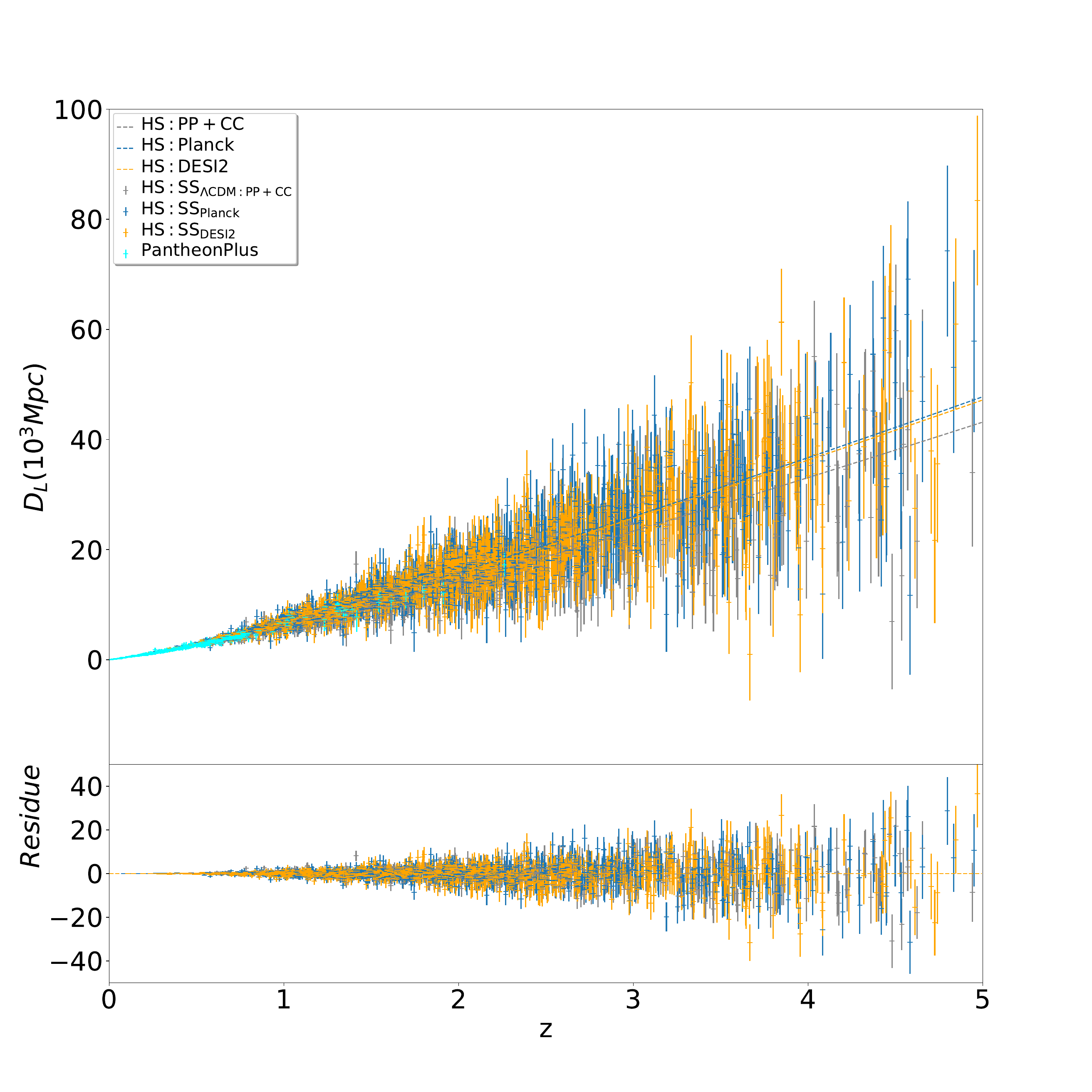}\\
	\includegraphics[width=8.65cm]{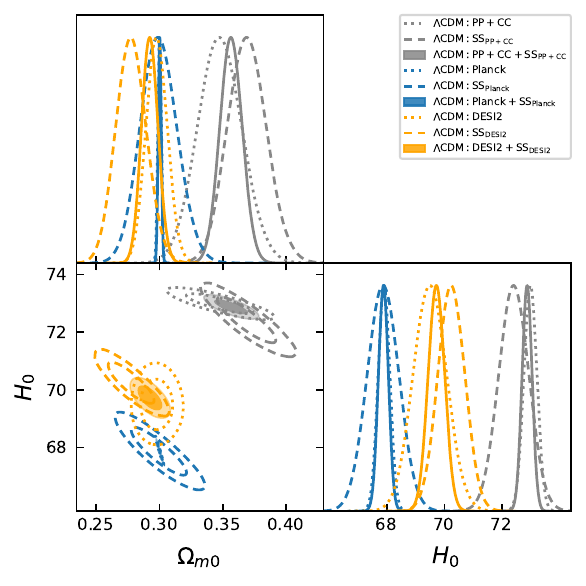}
	\includegraphics[width=8.65cm]{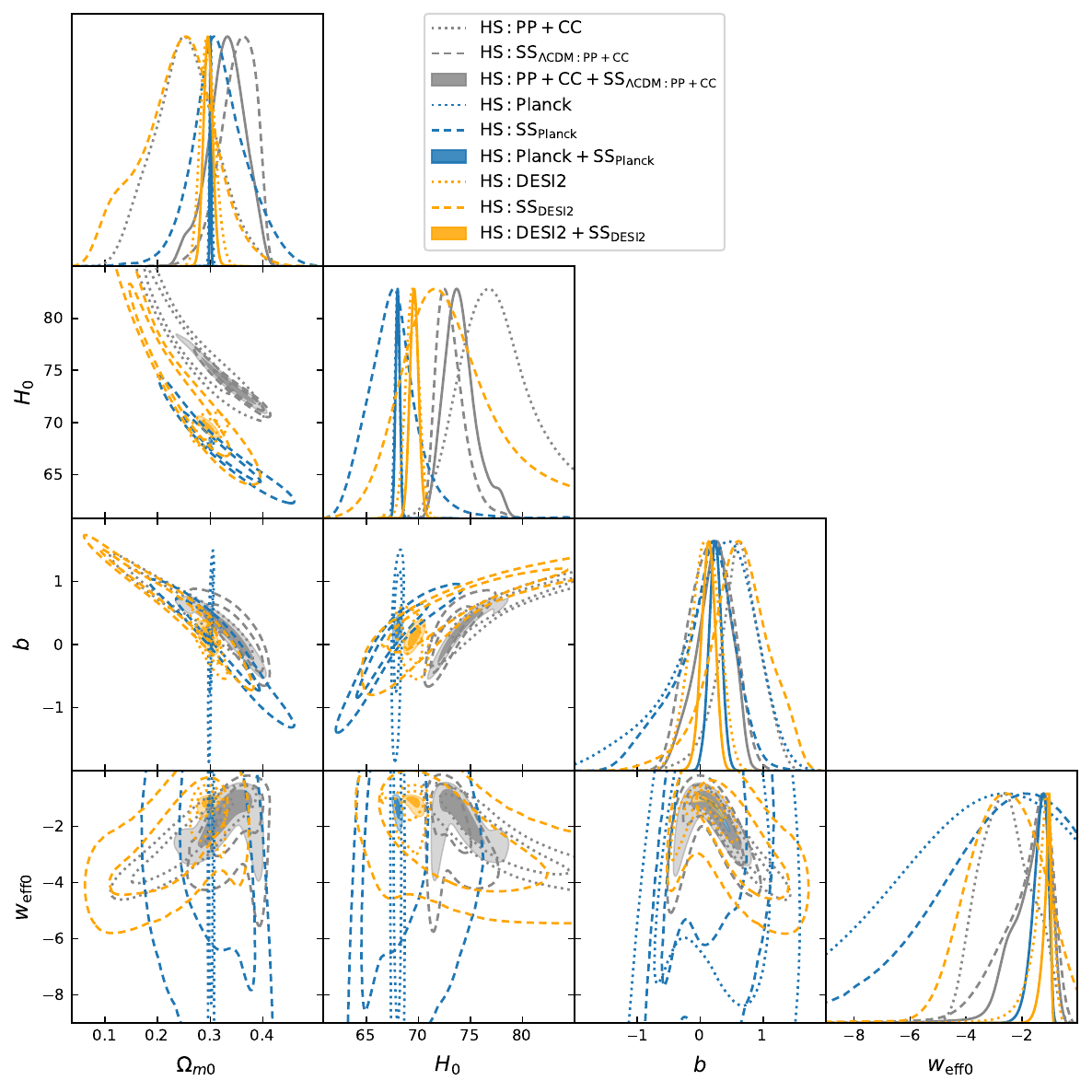}	
		\caption{
			\textit{Left:} Comparison between luminosity-distance measurements from real and
			simulated datasets. The cyan points with error bars correspond to the
			PantheonPlus sample (real SNe~Ia data). The gray/orange/blue points with error
			bars show the simulated standard siren datasets
			$\mathrm{SS}_{\rm DESI2}/\mathrm{SS}_{\rm PP+CC}/\mathrm{SS}_{\rm Planck}$,
			respectively. The gray/orange/blue solid curves represent the corresponding
			fiducial luminosity-distance relations used to generate the mock catalogs.
			\textit{Right:} One and two dimensional marginalized posteriors (with
			$1\sigma$ and $2\sigma$ confidence regions) for the parameters of $\Lambda$CDM
			and the Hu-Sawicki model. The simulated datasets
			$\Lambda$CDM:$\mathrm{SS}_{\rm DESI2}$/$\Lambda$CDM:$\mathrm{SS}_{\rm PP+CC}$/$\Lambda$CDM:$\mathrm{SS}_{\rm Planck}$
			are generated assuming $\Lambda$CDM with the EM best-fit fiducial parameters
			$(\Omega_{m0}^{\rm fid},H_0^{\rm fid})=(0.348,72.99)/(0.298,69.50)/(0.300,67.89)$.
			For the Hu-Sawicki simulations, we adopt
			$(\Omega_{m0}^{\rm fid},H_0^{\rm fid},b^{\rm fid})=(0.348,72.99,0)$,
			$(0.296,69.37,-0.06\times 10^{-3})$, and
			$(0.301,67.95,0.15\times 10^{-3})$
			for HS:$\mathrm{SS}_{\rm PP+CC}$, HS:$\mathrm{SS}_{\rm DESI2}$, and
			HS:$\mathrm{SS}_{\rm Planck}$, respectively.}
	\label{fr1tri}        
\end{figure*}
			
		\begin{table*}
			\begin{center}
                \caption{Constraints on the primary parameters $(\Omega_{m0},\,H_0,\,b)$ and the
                	derived quantities $(w_{\rm eff0},\,\text{Hubble tension},\,\chi^2)$ for
                	$\Lambda$CDM, the Hu-Sawicki model and the Starobinsky model. For the
                	standard siren simulations, the best fitted values inferred from the EM datasets
                	(PP+CC/Planck/DESI2) are adopted as fiducial (baseline) parameters for the
                	corresponding mock catalogs. In particular, HS:$\mathrm{SS}_{\Lambda{\rm CDM:PP+CC}}$
                	is generated using $(\Omega_{m0},\,H_0,\,b)=(0.348,\,72.99,\,0)$ which is the
                	best-fit parameters of the $\Lambda$CDM model. The Hubble tension is quantified
                	with respect to the SH0ES determination
                	$H_0 = 73.17 \pm 0.86~\mathrm{km\,s^{-1}\,Mpc^{-1}}$
                	\cite{Riess:2021jrx,Breuval:2024lsv}, as described in
                	Sec.~\ref{method}.}
				\begin{tabular}{|l|ccc|c|c|c|}
				\hline   
				Data&$\Omega_{m0}$& $H_0(\rm km/s/Mpc)$ & $b$  & $w_{\rm eff0}$&Hubble Tension & $\chi^2$\\  
				\hline   
				\hline
				$\Lambda$CDM:PP+CC& $0.348^{+0.017+0.033}_{-0.017-0.033}$  & $72.99^{+0.21+0.42}_{-0.22-0.43}$  &$0$&$-1$  & $0.20\sigma$&  $1774.6$\\
				
				$\Lambda$CDM:SS$_{\rm PP+CC}$&$0.369^{+0.015+0.03}_{-0.014-0.028}$  &$72.41^{+0.50+0.99}_{-0.50-0.98}$ &$0$&$-1$  &  $0.76\sigma$ &  $981.8$ \\
				
				$\Lambda$CDM:PP+CC+SS$_{\rm PP+CC}$&$0.357^{+0.008+0.017}_{-0.009-0.017}$  &$72.87^{+0.16+0.33}_{-0.16-0.32}$  &$0$&$-1$&  $0.34\sigma$&  $2755.6$  \\

				\hline
				$\Lambda$CDM:	   Planck & $0.300^{+0.001+0.003}_{-0.001-0.002}$  &$67.89^{+0.20+0.38}_{-0.19-0.38}$ &$0$&$-1$     &  $5.99\sigma$ &  $2.8$ \\	
				
				$\Lambda$CDM:SS$_{\rm    Planck}$ & $0.300^{+0.013+0.028}_{-0.014-0.027}$     &$ 67.86^{+0.53+1.04}_{-0.52-1.03}$   &$0$&$-1$     & $5.27\sigma$&  $989.7$ \\
				
				$\Lambda$CDM:   Planck+SS$_{\rm    Planck}$& $0.300^{+0.001+0.002}_{-0.001-0.002}$     &$ 67.87^{+0.17+0.33}_{-0.16-0.32}$   &$0$&$-1$    & $6.06\sigma$&  $990.6$  \\
				
				\hline
				
				$\Lambda$CDM:DESI2&  $0.298^{+0.008+0.016}_{-0.008-0.016}$   &$69.50^{+0.55+1.10}_{-0.55-1.10}$     &$0$&$-1$ &$3.60\sigma$ &  $16.8$   \\
				
				$\Lambda$CDM:SS$_{\rm DESI2}$&$0.278^{+0.011+0.023}_{-0.012-0.023}$  & $70.24^{+0.45+0.89}_{-0.45-0.89}$  &$0$&$-1$     &$3.02\sigma$ &  $987.3$  \\
				
				$\Lambda$CDM:DESI2+SS$_{\rm DESI2}$&$0.292^{+0.007+0.013}_{-0.006-0.012}$  & $69.72^{+0.28+0.54}_{-0.28-0.54}$   &$0$&$-1$      &$3.82\sigma$ &  $1004.3$   \\
				\hline
				\hline
				HS:	PP+CC&$0.252^{+0.058+0.109}_{-0.056-0.108}$      & $78.65^{+1.88+9.73}_{-5.34-7.06}$   &$0.62^{+0.33+0.64}_{-0.29-0.95}$   &$-2.628^{+1.608+1.644}_{-1.067-1.399}$ & $1.48\sigma$ &  $1771.8$  \\      
				
				HS:SS$_{\rm \Lambda CDM:PP+CC}$&  $0.351^{+0.045+0.049}_{-0.016-0.057}$       & $73.16^{+0.63+2.71}_{-1.65-2.10}$    &$0.17^{+0.35+0.58}_{-0.32-0.60}$     &$-1.971^{+1.038+1.090}_{-0.296-1.921}$ & $0.01\sigma$ &  $982.3$ \\
				
				HS:PP+CC+SS$_{\rm \Lambda CDM:PP+CC}$&$0.331^{+0.036+0.069}_{-0.031-0.359}$  &$74.04^{+1.00+3.93}_{-1.74-3.04}$ &$0.21^{+0.32+0.47}_{-0.21-0.53}$   & $-1.743^{+0.799+0.872}_{-0.265-1.220}$ &   $0.54\sigma$ &  $2756.0$  \\

				\hline
				HS:   Planck &$0.301^{+0.002+0.004}_{-0.002-0.004}$     &$67.95^{+0.22+0.44}_{-0.22-0.44}$   &$0.15^{+0.80+1.09}_{-0.37-1.45}$   & $-5.584^{+4.778+5.075}_{-1.554-17.189}$&$5.88\sigma$	&  $4.7$ \\	
				
				HS:SS$_{\rm    Planck}$ & $0.319^{+0.044+0.108}_{-0.052-0.098}$     & $67.60^{+1.71+4.45}_{-2.51-4.67}$  & $0.04^{+0.57+0.77}_{-0.32-1.08}$   & $-4.471^{+3.729+4.318}_{-1.804-17.575}$   &$2.44\sigma$ &  $987.4$ \\
				
				HS:   Planck+SS$_{\rm    Planck}$& $0.300^{+0.002+0.003}_{-0.001-0.002}$     &$ 67.99^{+0.17+0.35}_{-0.19-0.36}$   & $0.24^{+0.09+0.18}_{-0.10-0.19}$     &$-1.383^{+0.335+0.412}_{-0.140-0.530}$ & $5.90\sigma$ & $992.5$  \\  
				
					\hline
				HS:DESI2& $0.296^{+0.012+0.028}_{-0.015-0.025}$          & $69.37^{+0.62+1.21}_{-0.60-1.26}$    & $0.06^{+0.26+0.42}_{-0.19-0.45}$    &$-1.381^{+0.410+0.430}_{-0.027-0.982}$ & $3.60\sigma$&  $17.7$ \\
				
				HS:SS$_{\rm DESI2}$&$0.235^{+0.077+0.109}_{-0.054-0.142}$   & $75.06^{+1.69+16.43}_{-7.27-9.45}$     & $0.60^{+0.47+0.93}_{-0.43-0.88}$     	& $-2.845^{+1.429+1.890}_{-0.920-1.358}$ &$0.41\sigma$&  $984.9$\\
				
				HS:DESI2+SS$_{\rm DESI2}$&$0.296^{+0.008+0.016}_{-0.008-0.016}$  & $69.57^{+0.40+0.80}_{-0.39-0.79}$     &  $0.13^{+0.11+0.19}_{-0.10-0.22}$   &$-1.152^{+0.164+0.175}_{-0.045-0.332} $ &$3.80\sigma$ &  $1002.0$  \\

				\hline
				\hline
				Star\Romannum{1}:PP+CC&$0.293^{+0.045+0.075}_{-0.029-0.083}$      & $73.05^{+0.49+3.01}_{-0.92-1.82}$   &$-1.05^{+0.13+1.05}_{-0.28-0.33}$   &$-0.837^{+0.057+0.100}_{-0.044-0.161}$ & $0.11\sigma$ &  $1770.2$ \\  
				
				Star\Romannum{2}:PP+CC&$0.293^{+0.045+0.076}_{-0.029-0.086}$      & $73.08^{+0.47+3.36}_{-0.95-1.85}$   &$1.05^{+0.29+0.34}_{-0.13-1.05}$  &$-0.837^{+0.056+0.100}_{-0.044-0.161}$ & $0.08\sigma$  &  $1770.2$  \\         
				
				Star\Romannum{1}:SS$_{\rm PP+CC}$&  $0.306^{+0.014+0.084}_{-0.044-0.056}$       & $71.80^{+3.46+3.57}_{-1.13-5.81}$    &$-1.17^{+0.14+1.17}_{-0.47-0.45}$   &$-0.723^{+0.059+0.483}_{-0.265-0.299}$& $0.55\sigma$ &  $983.6$   \\  
				
				Star\Romannum{2}:SS$_{\rm PP+CC}$&  $0.306^{+0.019+0.062}_{-0.031-0.051}$       & $72.55^{+2.61+3.27}_{-1.11-4.50}$    &$1.00^{+0.46+0.46}_{-0.17-1.00}$   &$-0.803^{+0.065+0.314}_{-0.184-0.227}$ & $0.30\sigma$ &  $988.6$\\    
				
				Star\Romannum{1}:PP+CC+SS$_{\rm PP+CC}$&$0.297^{+0.013+0.028}_{-0.015-0.030}$  &$72.48^{+0.67+1.06}_{-0.37-1.16}$ &$-1.16^{+0.08+0.24}_{-0.13-0.22}$  &  
				$-0.795^{+0.036+0.090}_{-0.041-0.086}$ &  $0.69\sigma$ &  $2752.6$  \\ 
		
				Star\Romannum{2}:PP+CC+SS$_{\rm PP+CC}$&$0.304^{+0.013+0.027}_{-0.013-0.028}$  &$72.47^{+0.60+0.94}_{-0.37-1.05}$ & $1.10^{+0.15+0.19}_{-0.09-1.10}$   	& $-0.810^{+0.044+0.090}_{-0.040-0.100}$ &  $0.71\sigma$&  $2756.1$                \\

				\hline
				Star\Romannum{1}:   Planck &$0.301^{+0.002+0.004}_{-0.002-0.003}$     &$67.97^{+0.21+0.42}_{-0.21-0.40}$   &$-0.53^{+0.30+0.53}_{-0.34-0.45}$   & $-0.950^{+0.019+0.097}_{-0.053-0.057}$ &$5.87\sigma$ &  $2.5$ \\ 
				
				Star\Romannum{2}:   Planck &$0.301^{+0.002+0.004}_{-0.002-0.004}$     &$67.97^{+0.21+0.41}_{-0.21-0.40}$   &$0.53^{+0.16+0.45}_{-0.53-0.53}$  & $-0.950^{+0.019+0.097}_{-0.054-0.057}$ &$5.87\sigma$&  $2.5$  \\
				
				Star\Romannum{1}:SS$_{\rm    Planck}$ & $0.335^{+0.012+0.081}_{-0.034-0.052}$     & $65.74^{+2.37+2.68}_{-0.28-5.16}$  & $-0.73^{+0.73+0.73}_{-0.28-0.75}$   & $-0.839^{+0.046+0.653}_{-0.205-0.238}$ &$4.70\sigma$ &  $984.4$ \\

				Star\Romannum{2}:SS$_{\rm    Planck}$ & $0.318^{+0.007+0.119}_{-0.031-0.054}$     & $66.99^{+1.89+2.72}_{-0.08-8.60}$  & $0.54^{+0.21+1.02}_{-0.54-0.54}$    & $-0.898^{+0.006+0.877}_{-0.139-0.196}$ &$4.73\sigma$ &  $988.8$ \\ 
				
				Star\Romannum{1}:   Planck+SS$_{\rm    Planck}$& $0.301^{+0.001+0.003}_{-0.001-0.003}$     &$ 67.87^{+0.17+0.33}_{-0.17-0.34}$   & $-0.40^{+0.40+0.40}_{-0.12-0.29}$  &$-0.974^{+0.011+0.041}_{-0.027-0.030}$& $6.05\sigma$ &  $987.1$   \\	     
				
				Star\Romannum{2}:   Planck+SS$_{\rm    Planck}$& $0.301^{+0.001+0.003}_{-0.002-0.003}$     &$ 68.00^{+0.16+0.33}_{-0.18-0.36}$   & $0.51^{+0.26+0.27}_{-0.12-0.51}$   &$-0.960^{+0.019+0.049}_{-0.036-0.043}$ & $5.90\sigma$ &  $989.7$\\  
				\hline
				Star\Romannum{1}:DESI2& $0.325^{+0.018+0.042}_{-0.030-0.039}$          & $64.32^{+5.16+5.55}_{-4.09-4.73}$    & $-0.99^{+0.11+0.99}_{-0.43-0.40}$  &$-0.778^{+0.135+0.271}_{-0.215-0.247}$ & $1.88\sigma$&  $14.4$\\

				Star\Romannum{2}:DESI2& $0.325^{+0.018+0.042}_{-0.030-0.039}$          & $64.32^{+5.16+5.55}_{-4.09-4.74}$    & $0.99^{+0.43+0.40}_{-0.12-0.99}$   &$-0.778^{+0.134+0.273}_{-0.213-0.248}$& $1.88\sigma$&  $14.4$ \\
				
				Star\Romannum{1}:SS$_{\rm DESI2}$&$0.335^{+0.016+0.030}_{-0.010-0.028}$   & $64.88^{+0.28+0.64}_{-0.34-1.02}$     & $-0.51^{+0.51+0.51}_{-0.19-0.52}$   & $-0.938^{+0.024+0.132}_{-0.069-0.072}$&$9.07\sigma$&  $983.2$  \\

				Star\Romannum{2}:SS$_{\rm DESI2}$&$0.358^{+0.022+0.053}_{-0.030-0.046}$   & $62.80^{+2.05+2.62}_{-1.24-3.06}$     & $0.96^{+0.43+0.42}_{-0.14-0.96}$    & $-0.738^{+0.108+0.367}_{-0.252-0.291}$&$5.59\sigma$&  $987.4$  \\  
				
				Star\Romannum{1}:DESI2+SS$_{\rm DESI2}$&$0.334^{+0.014+0.033}_{-0.019-0.031}$  & $63.75^{+1.23+2.05}_{-1.02-2.32}$     &  $-1.10^{+0.10+0.23}_{-0.13-0.23}$  &$-0.753^{+0.058+0.183}_{-0.101-0.150} $ &$6.65\sigma$ &  $1000.2$  \\

				Star\Romannum{2}:DESI2+SS$_{\rm DESI2}$&$0.340^{+0.015+0.032}_{-0.017-0.031}$  & $62.72^{+1.01+2.09}_{-1.07-2.05}$     &  $1.19^{+0.11+0.19}_{-0.08-0.19}$   &$-0.696^{+0.068+0.196}_{-0.098-0.151} $ 	&$7.74\sigma$  &  $1000.5$ \\

				\hline
			\end{tabular}
		\label{tfr}
	\end{center}
\end{table*} 
\begin{figure*}[!htb]
	\centering
	\includegraphics[width=8.65cm]{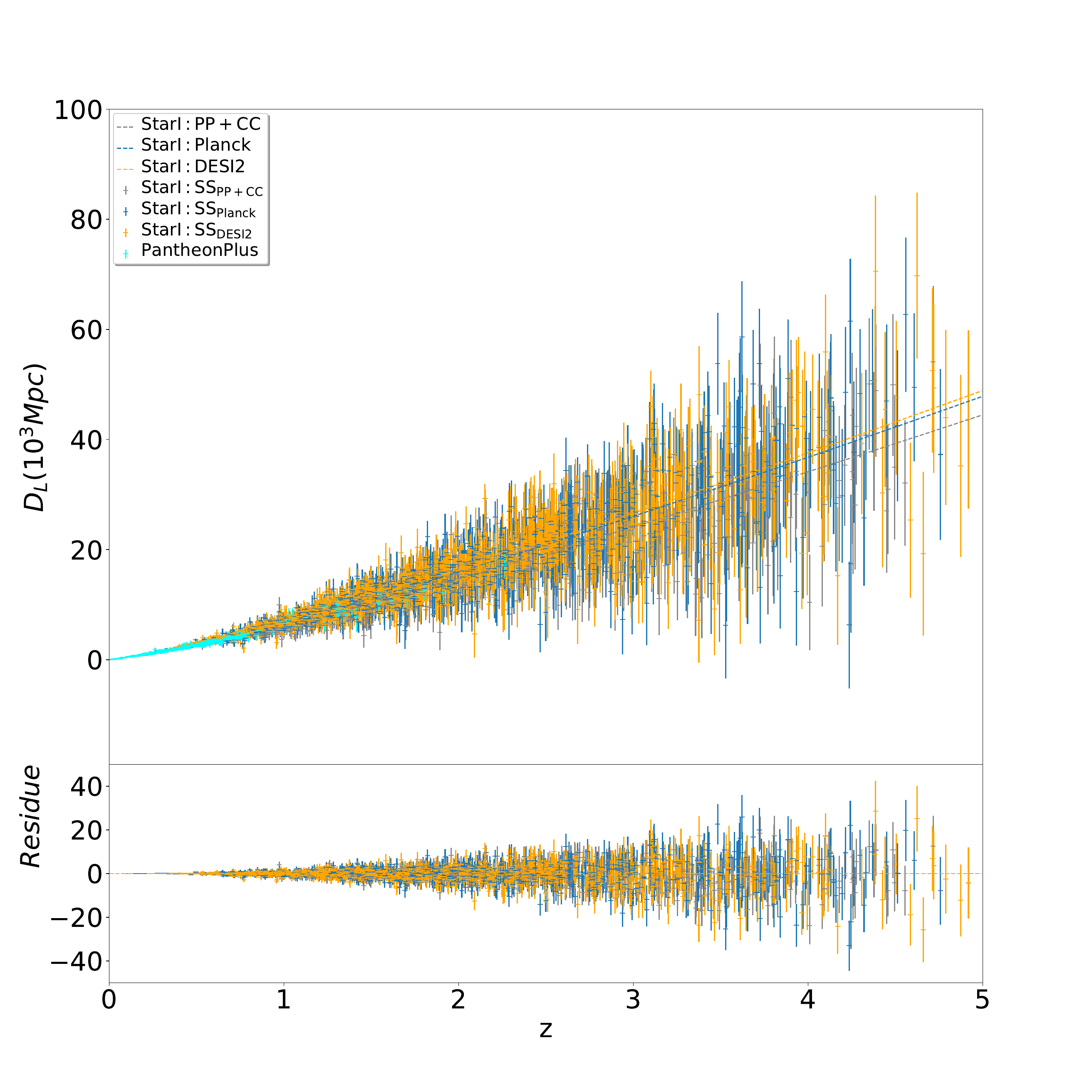}
		\includegraphics[width=8.65cm]{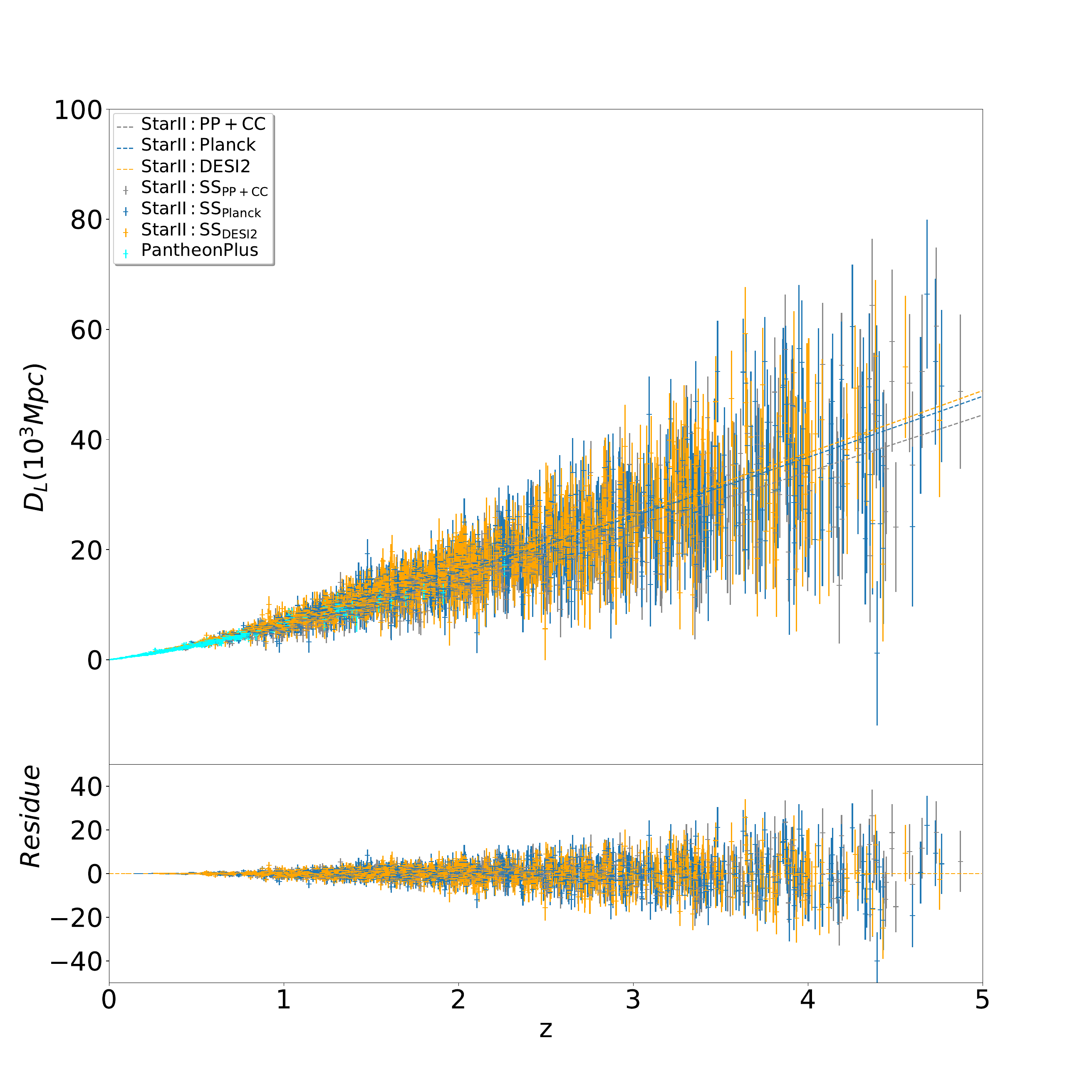}\\
	\includegraphics[width=8.65cm]{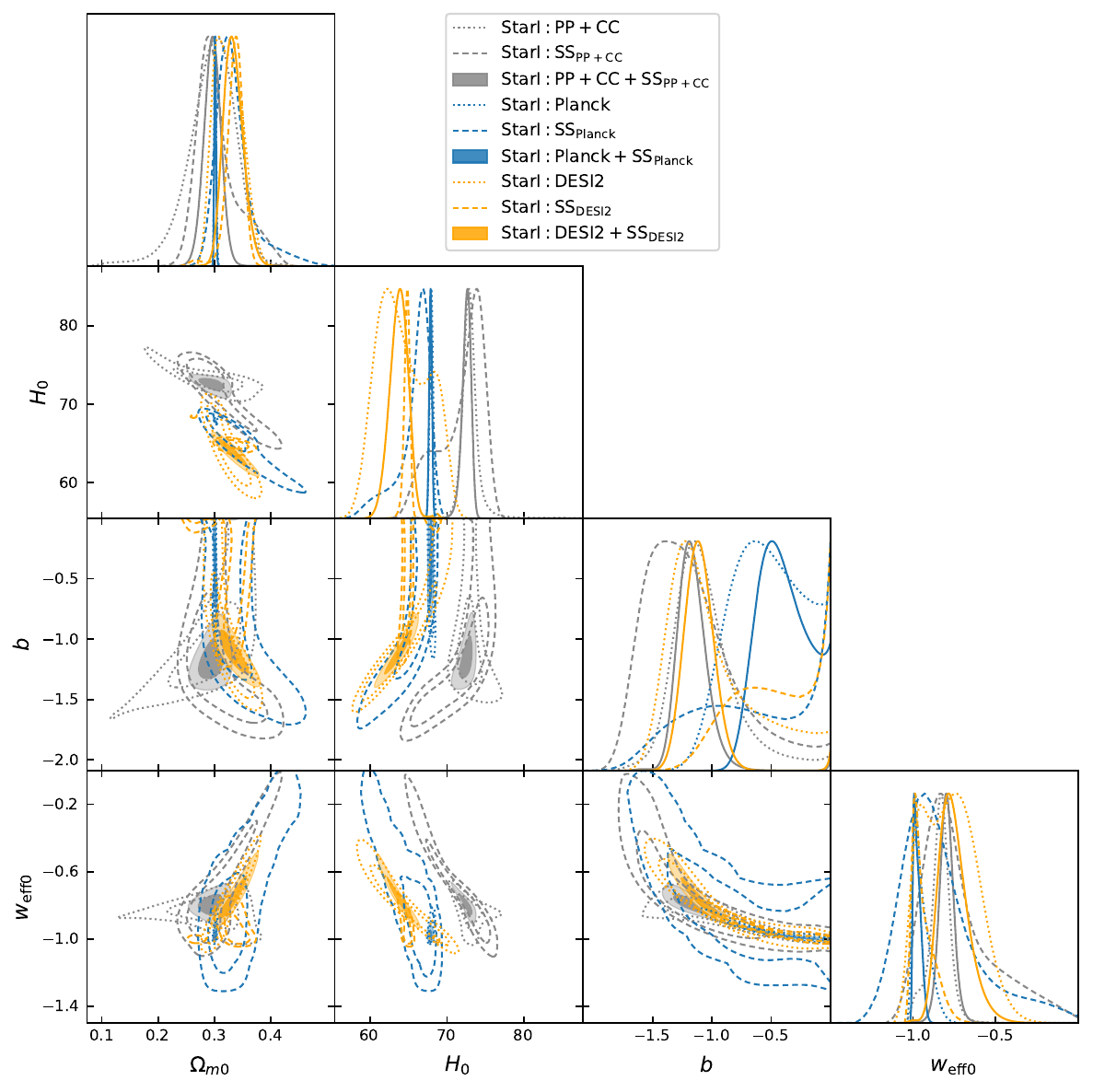}	
				\includegraphics[width=8.65cm]{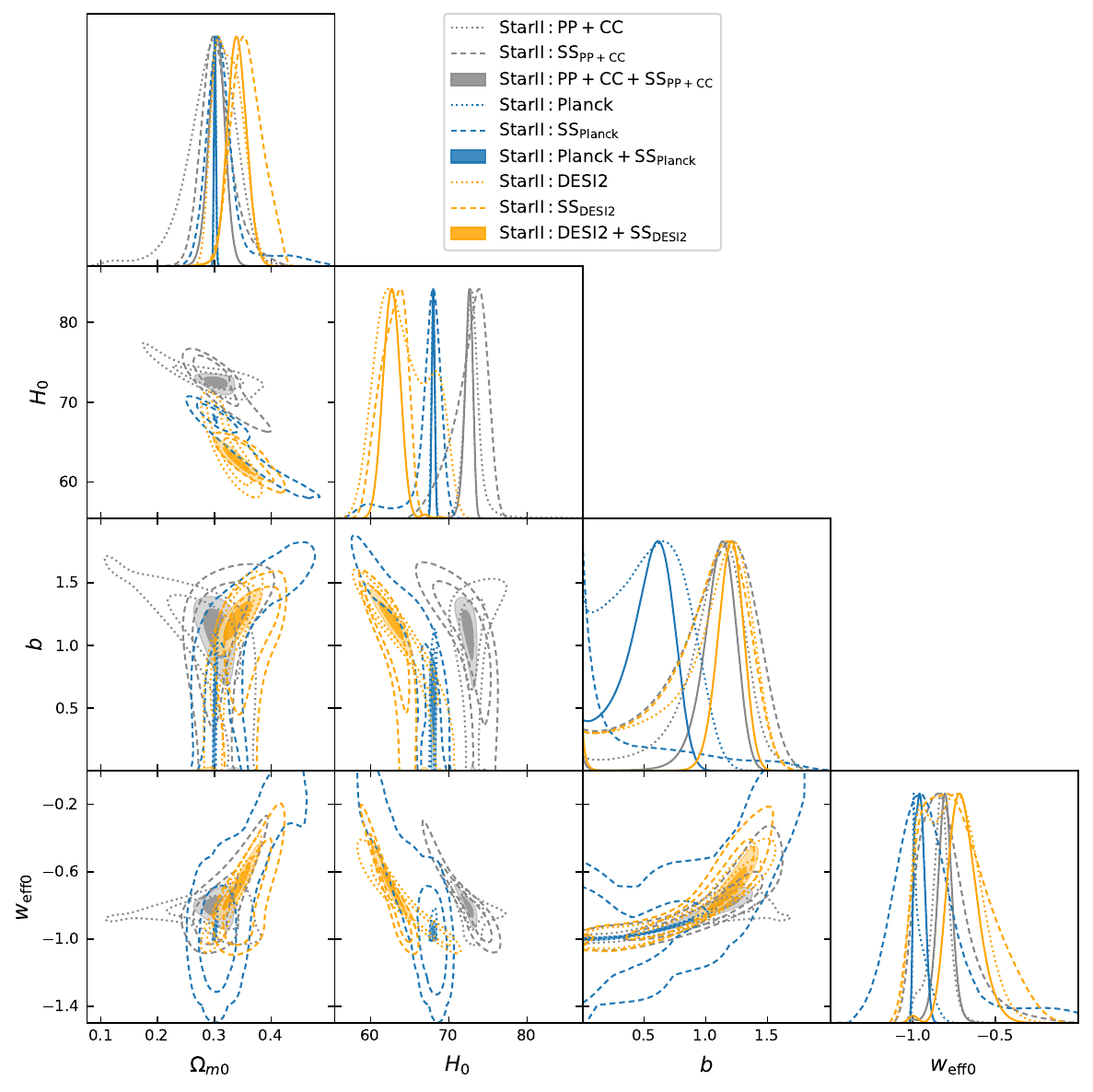}	\\
	\caption{Same as Fig.~\ref{fr1tri}, but for the Starobinsky\Romannum{1} ($b<0$)
		and Starobinsky\Romannum{2} ($b>0$) models. For the corresponding standard siren
		simulations, the EM best-fit values from PP+CC/Planck/DESI2 are adopted as
		fiducial (baseline) parameters. Specifically, the mock catalogs
		Star\Romannum{1}:$\mathrm{SS}_{\rm PP+CC}$, Star\Romannum{1}:$\mathrm{SS}_{\rm DESI2}$,
		and Star\Romannum{1}:$\mathrm{SS}_{\rm Planck}$ are generated using
		$(\Omega_{m0}^{\rm fid},H_0^{\rm fid},b^{\rm fid})=(0.293,73.05,-1.05\times10^{-3})$,
		$(0.325,64.32,-0.99\times10^{-3})$, and $(0.301,67.97,-0.53\times10^{-3})$,
		respectively. Likewise, the mock catalogs
		Star\Romannum{2}:$\mathrm{SS}_{\rm PP+CC}$, Star\Romannum{2}:$\mathrm{SS}_{\rm DESI2}$,
		and Star\Romannum{2}:$\mathrm{SS}_{\rm Planck}$ are generated using
		$(\Omega_{m0}^{\rm fid},H_0^{\rm fid},b^{\rm fid})=(0.293,73.08,1.05\times10^{-3})$,
		$(0.325,64.32,0.99\times10^{-3})$, and $(0.301,67.97,0.53\times10^{-3})$,
		respectively.}
	\label{fr2tri}        
\end{figure*}
\begin{table*}
		\begin{center}
    \caption{Constraints on the derived quantities $(F_{R0},\,F_{RR0},\,r_0,\,m_0)$
    	and the information criteria $(\Delta{\rm AIC},\,\Delta{\rm BIC})$ for the
    	Hu-Sawicki and Starobinsky models. Since HS:$\mathrm{SS}_{\Lambda{\rm CDM:PP+CC}}$
    	is generated using the $\Lambda$CDM best-fit fiducial parameters
    	$(\Omega_{m0},\,H_0,\,b)=(0.348,\,72.99,\,0)$, we also report the corresponding
    	AIC and BIC comparison between HS:$\mathrm{SS}_{\Lambda{\rm CDM:PP+CC}}$ and
    	$\Lambda$CDM:$\mathrm{SS}_{\rm PP+CC}$.}

		\begin{tabular}{|l|cc|cc|cc|}
			\hline   
			Data &	$F_{R0}(10^{-3})$ &  $F_{RR0}(10^{-7})$  & $r_0$ & $m_0(10^{-3})$   & $\Delta $AIC & $\Delta$BIC 	\\ 
			\hline 
			\hline

		HS:PP+CC& $-47.72^{+14.21+47.45}_{-26.10-41.00}$    & $12.75^{+4.15+4.71}_{-0.42-11.49}$  &$-1.602^{+0.104+0.166}_{-0.055-0.182}$ &$86.01^{+48.78+62.60}_{-16.09-80.55}$   &  $-0.8$ &  $4.7$  \\
		
		HS:SS$_{\rm \Lambda CDM:PP+CC}$  &  $-12.00^{+17.82+56.03}_{-37.58-47.58}$    &  $3.51^{+14.49+15.64}_{-59.29-23.45}$&$-1.726^{+0.156+0.203}_{-0.077-0.244}$    &$21.32^{+71.12+88.35}_{-10.89-67.87}$   &  $2.5$ &  $7.4$  \\
		
		HS:PP+CC+SS$_{\rm \Lambda CDM:PP+CC}$ &  $-16.62^{+14.18+50.51}_{-29.22-39.39}$    &  $5.30^{+10.36+12.46}_{-2.94-20.89}$&$-1.716^{+0.130+0.155}_{-0.046-0.218}$  &$30.80^{+56.21+73.29}_{-38.72-38.24}$  &  $2.4$ &  $8.3$ \\

		\hline
		HS:   Planck &  $10.46^{+1.30+215.33}_{-84.51-96.43}$    &$-13.53^{+41.79+45.24}_{-10.09-130.32}$& $-1.955^{+0.422+67.129}_{-0.219-28.968}$  &$-32.53^{+163.31+174.86}_{-19.69-467.53}$  &  $ 3.9$ &  $3.0$  \\	 
		
		HS:SS$_{\rm    Planck}$ & $4.35^{+16.93+125.93}_{-60.57-77.75}$      & $-8.35^{+29.39+32.58}_{-0.36-98.91}$  &$-1.830^{+0.267+0.338}_{-0.065-0.570}$    &$-14.56^{+121.36+148.31}_{-23.75-265.                                                                                                                                                                                                                    68}$  &  $-$ &  $-$  \\
		
		HS:   Planck+SS$_{\rm    Planck}$ & $-21.46^{+7.04+15.86}_{-8.49-15.38}$    & $9.40^{+3.66+6.22}_{-2.73-9.04}$&$-1.714^{+0.044+0.078}_{-0.035-0.080}$     &$41.45^{+16.30+28.19}_{-12.73-39.76}$  &  $-$ &  $-$   \\
		
				\hline
		HS:DESI2& $-4.24^{+14.59+46.88}_{-27.24-39.00}$
		&$1.31^{+12.22+16.53}_{-5.32-21.51}$	&$-1.805^{+0.134+0.190}_{-0.068-0.228}$       & $7.15^{+54.69+75.27}_{-26.13-94.49}$  &  $2.9$ &  $3.4$  \\
		
		HS:SS$_{\rm DESI2}$& $-43.90^{+14.23+20.44}_{-41.74-54.21}$ &  $10.95^{+8.24+9.28}_{-1.96-25.33}$  &$-1.628^{+0.175+0.235}_{-0.066-0.312}$  &     $76.78^{+77.62+81.20}_{-9.39-132.6}$  &  $-$ &  $-$  \\
		
		HS:DESI2+SS$_{\rm DESI2}$& $-11.64^{+11.73+19.53}_{-10.90-18.28}$  &  $4.87^{+4.77+7.39}_{-3.02-8.21}$& $-1.767^{+0.053+0.089}_{-0.039-0.094}$      & $22.67^{+21.81+35.13}_{-22.80-38.08}$    &  $-$ &  $-$   \\	
			
			\hline  
			\hline
		Star\Romannum{1}:PP+CC & $-0.92^{+0.28+0.57}_{-0.28-0.56}$    & $0.57^{+0.18+0.40}_{-0.21-0.39}$ &$-1.824^{+0.031+0.032}_{-0.022-0.028}$   & $2.77^{+0.84+1.67}_{-0.84-1.72}$    &  $ - 2.4$ &  $3.1$   \\
		
		Star\Romannum{2}:PP+CC  	&$-0.92^{+0.28+0.60}_{-0.28-0.56}$    & $0.57^{+0.19+0.40}_{-0.21-0.41}$     &$-1.825^{+0.032+0.055}_{-0.022-0.059}$ & $2.76^{+0.85+1.68}_{-0.91-1.79}$  &  $- 2.4$ &  $3.1$ \\  
		
		Star\Romannum{1}:SS$_{\rm PP+CC}$& $-1.01^{+0.35+0.43}_{-0.07-0.75}$      & $0.67^{+0.02+0.73}_{-0.32-0.37}$  &$-1.816^{+0.010+0.063}_{-0.031-0.040}$    &$3.04^{+0.20+2.24}_{-1.06-1.31}$  &  $-$ &  $-$ \\
		
		Star\Romannum{2}:SS$_{\rm PP+CC}$ &  $-1.01^{+0.27+0.40}_{-0.11-0.52}$    &  $0.64^{+0.06+0.47}_{-0.23-0.32}$  &$-1.815^{+0.012+0.045}_{-0.023-0.037}$ &$3.02^{+0.33+0.15}_{-0.79-1.19}$   &  $-$ &  $-$ \\
		
		Star\Romannum{1}:PP+CC+SS$_{\rm PP+CC}$ & $-0.92^{+0.12+0.22}_{-0.09-0.21}$    &$0.56^{+0.07+1.6}_{-0.08-0.16}$& $-1.823^{+0.010+0.021}_{-0.010-0.020}$ & $2.75^{+0.29+0.64}_{-0.34-0.65} $  &  $-$ &  $-$  \\
		
		Star\Romannum{2}:PP+CC+SS$_{\rm PP+CC}$ &  $-0.97^{+0.10+0.20}_{-0.10-0.22}$   &  $0.60^{+0.07+0.17}_{-0.08-0.15}$  &$-1.817^{+0.010+0.019}_{-0.010-0.020}$    &$2.92^{+0.32+0.66}_{-0.32-0.62}$   &  $-$ &  $-$ \\

				\hline
			Star\Romannum{1}:   Planck	&  $-0.95^{+0.02+0.03}_{-0.01-0.03}$    &$0.66^{+0.01+0.03}_{-0.01-0.02}$ & $-1.819^{+0.001+0.002}_{-0.002-0.003}$ &$2.84^{+0.04+0.08}_{-0.04-0.07}$   &  $1.7$ &  $0.8$ \\

			Star\Romannum{2}:   Planck  &  $-0.95^{+0.02+0.03}_{-0.01-0.03}$    &$0.66^{+0.01+0.03}_{-0.01-0.02}$  & $-1.819^{+0.001+0.002}_{-0.002-0.003}$ &$2.84^{+0.04+0.08}_{-0.04-0.07}$   &  $1.7$ &  $0.8$  \\

			Star\Romannum{1}:SS$_{\rm    Planck}$&$-1.26^{+0.33+0.47}_{-0.07-0.85}$    &$1.01^{+0.02+1.04}_{-0.37-0.49}$ & $-1.794^{+0.008+0.063}_{-0.026-0.039}$  &$3.79^{+0.20+2.53}_{-1.00-1.42}  $   &  $-$ &  $-$  \\
			
			Star\Romannum{2}:SS$_{\rm    Planck}$ & $-1.12^{+0.29+0.46}_{-0.01-1.31}$     & $0.87^{+0.05+1.68}_{-0.32-0.46}$&$-1.806^{+0.004+0.093}_{-0.024-0.041}$    &$3.36^{+0.04+3.90}_{-0.87-1.37}$  &  $-$ &  $-$  \\
			
			Star\Romannum{1}:   Planck+SS$_{\rm    Planck}$  & $-0.95^{+0.02+0.02}_{-0.01-0.02}$     & $0.66^{+0.01+0.03}_{-0.01-0.02}$  &$-1.820^{+0.001+0.003}_{-0.001-0.001}$   &$2.84^{+0.03+0.07}_{-0.04-0.06}$   &  $-$ &  $-$  \\

			Star\Romannum{2}:   Planck+SS$_{\rm    Planck}$ & $-0.94^{+0.01+0.02}_{-0.02-0.03}$       & $0.66^{+0.01+0.02}_{-0.01-0.02}$&$-1.820^{+0.001+0.003}_{-0.001-0.002}$  &$2.83^{+0.04+0.08}_{-0.04-0.07}$ 	   &  $-$ &  $-$ \\

			\hline
			Star\Romannum{1}:DESI2  & $-1.15^{+0.32+0.32}_{-0.13-0.39}$    &$0.94^{+0.15+0.51}_{-0.36-0.40}$  &$-1.802^{+0.013+0.032}_{-0.022-0.028}$    & $3.45^{+0.40+1.16}_{-0.80-0.97}$   &  $-0.4$ &  $0.1$  \\
			
			Star\Romannum{2}:DESI2  & $-1.15^{+0.27+0.32}_{-0.13-0.39}$   &$0.94^{+0.15+0.51}_{-0.36-0.40}$ &$-1.802^{+0.013+0.032}_{-0.022-0.028}$    & $3.45^{+0.40+1.16}_{-0.80-0.97}$	   &  $-0.4$ &  $0.1$  \\
			
			Star\Romannum{1}:SS$_{\rm DESI2}$ & $-1.24^{+0.10+0.25}_{-0.14-0.27}$ &  $0.99^{+0.12+0.26}_{-0.12-0.24}$   & $-1.794^{+0.011+0.022}_{-0.008-0.021}$  & $3.72^{+0.40+0.82}_{-0.31-0.77}$  &  $-$ &  $-$  \\

			Star\Romannum{2}:SS$_{\rm DESI2}$& $-1.47^{+0.33+0.46}_{-0.19-0.58}$  &  $1.30^{+0.19+0.72}_{-0.42-0.55}$ &  $-1.777^{+0.017+0.043}_{-0.024-0.035}$    & $4.42^{+0.57+1.73}_{-1.01-1.39}$ 	  &  $-$ &  $-$ \\
			
			Star\Romannum{1}:DESI2+SS$_{\rm DESI2}$& $-1.22^{+0.18+0.27}_{-0.12-0.32}$    &  $1.01^{+0.12+0.37}_{-0.21-0.31}$ & $-1.796^{+0.011+0.026}_{-0.014-0.023}$    & $3.67^{+0.34+0.95}_{-0.54-0.83}$    &  $-$ &  $-$   \\

			Star\Romannum{2}:DESI2+SS$_{\rm DESI2}$ & $-1.28^{+0.16+0.28}_{-0.13-0.32}$&  $1.10^{+0.15+0.40}_{-0.21-0.30}$   & $-1.791^{+0.011+0.025}_{-0.013-0.023}$    & $3.84^{+0.40+0.97}_{-0.50-0.83}$   &  $-$ &  $-$   \\	
		\hline
		\end{tabular}
		
		\label{taic}
	\end{center}
\end{table*} 
\begin{figure*}[!htb]
	\centering	
	\includegraphics[width=5.41cm]{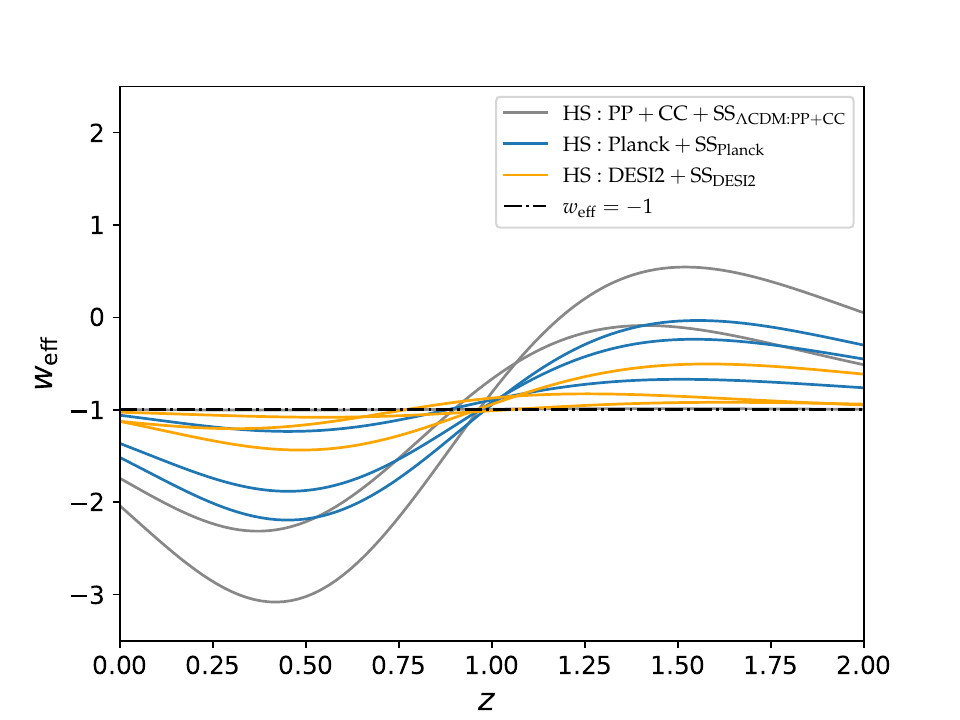}
	\includegraphics[width=5.41cm]{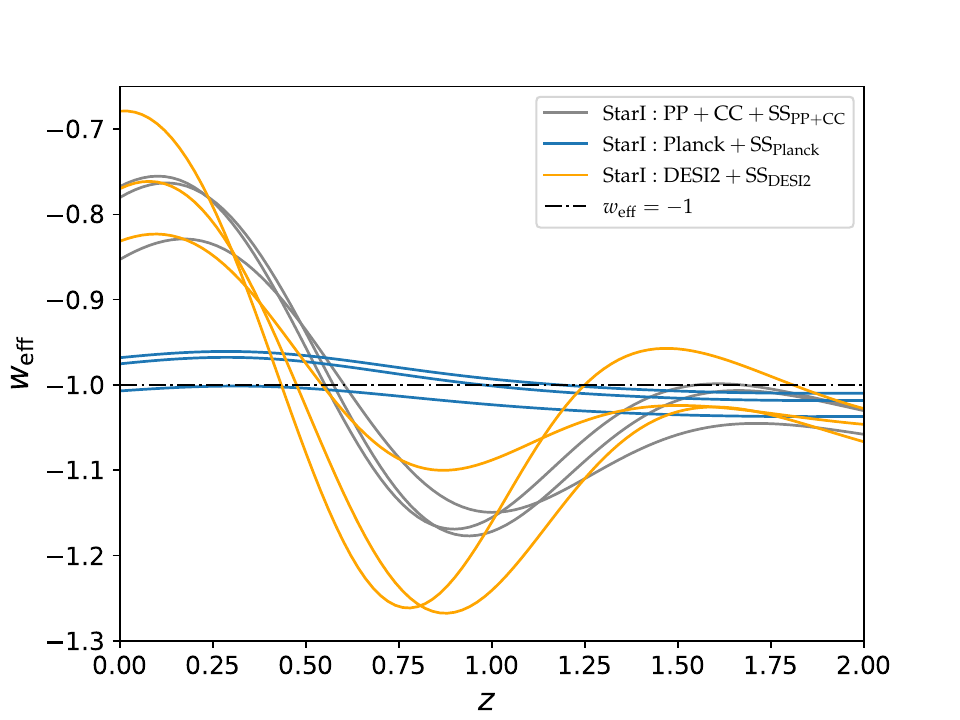}
	\includegraphics[width=5.41cm]{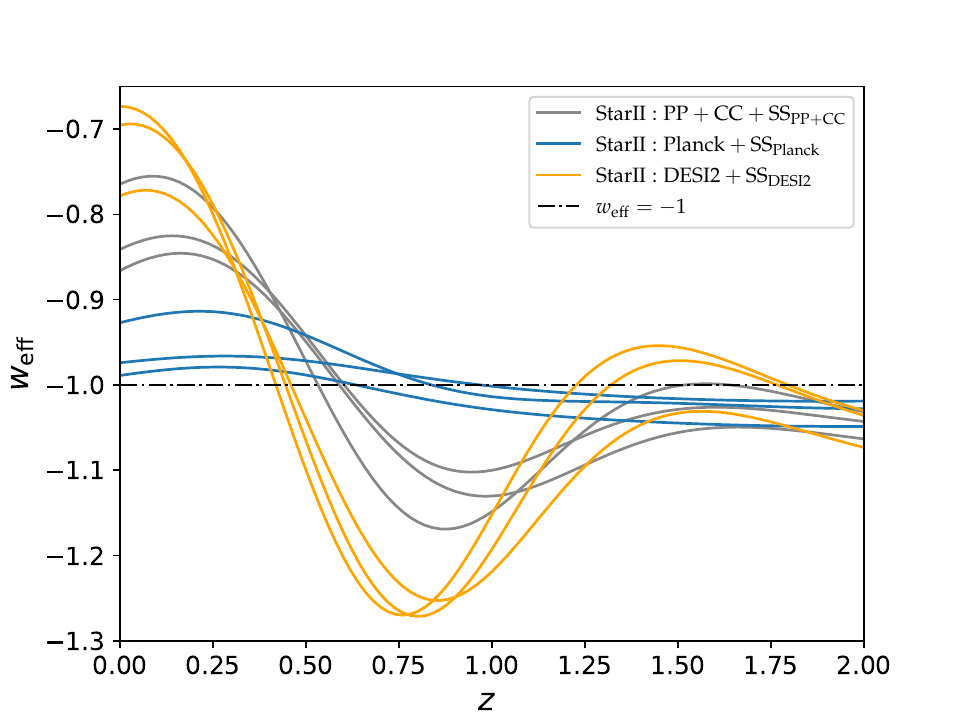}\\
	\caption{  The evolutions of   $w_{\rm eff}$   for the Hu-Sawicki (left),  Starobinsky\Romannum{1} (middle) and Starobinsky\Romannum{2} (right)  models, which are based on the the $1\sigma$ regimes of PP+CC+SS$_{\rm \Lambda CDM:PP+CC}$/PP+CC+SS$_{\rm PP+CC}$,  DESI2+SS$_{\rm DESI2}$ and  Planck+SS$_{\rm    Planck}$ constraining results.}
	\label{frw}        
\end{figure*}
\begin{figure*}[!htb]
	\centering	
	\includegraphics[width=4.1cm]{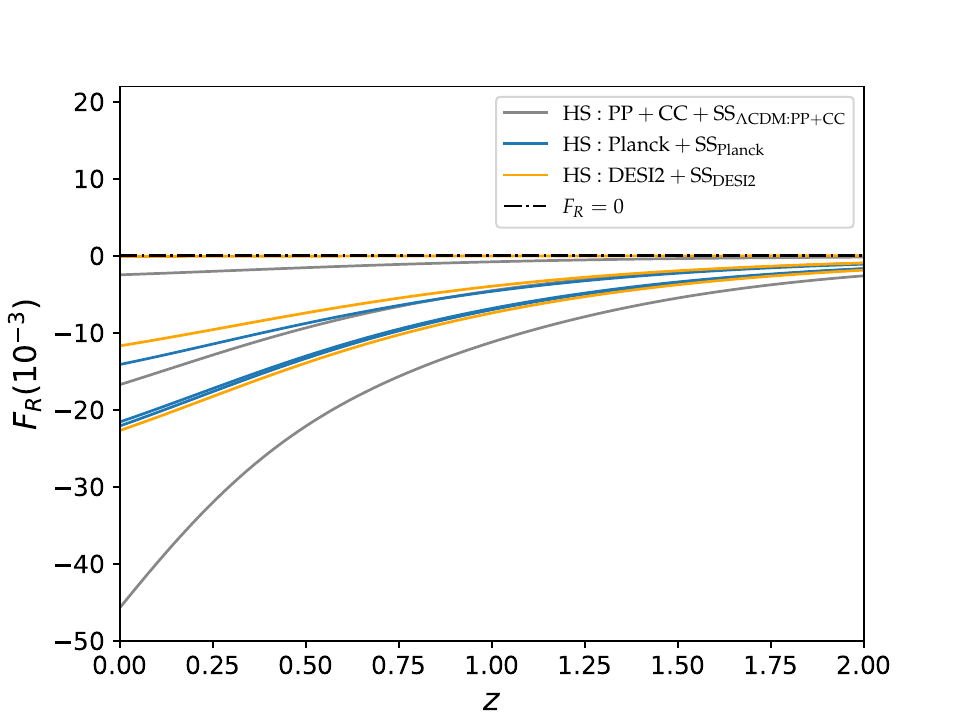}
	\includegraphics[width=4.1cm]{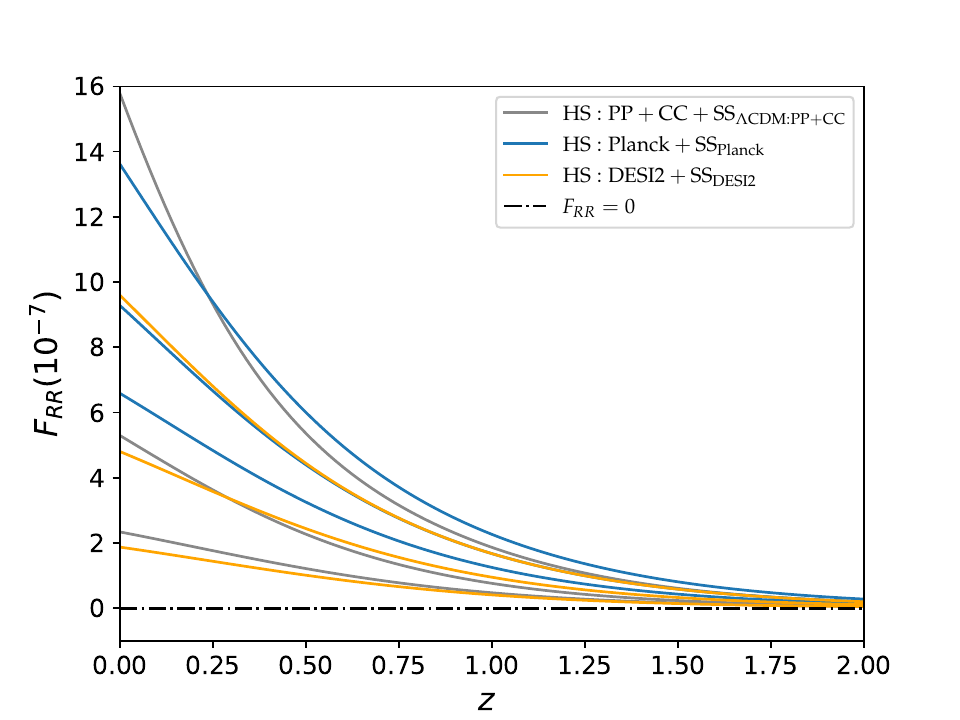}
	\includegraphics[width=4.1cm]{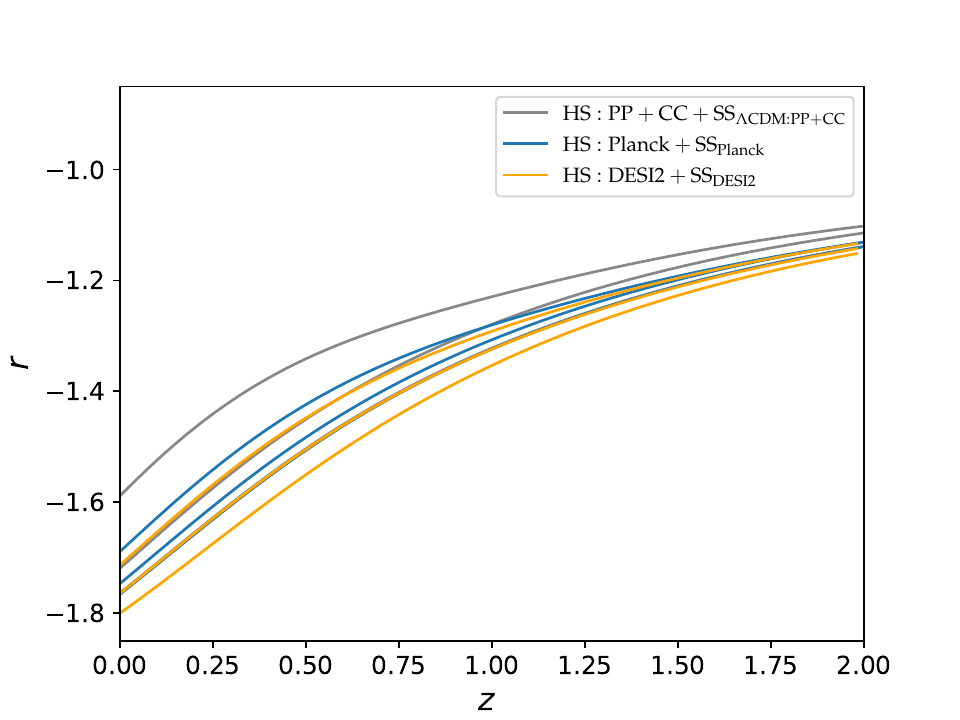}
	\includegraphics[width=4.1cm]{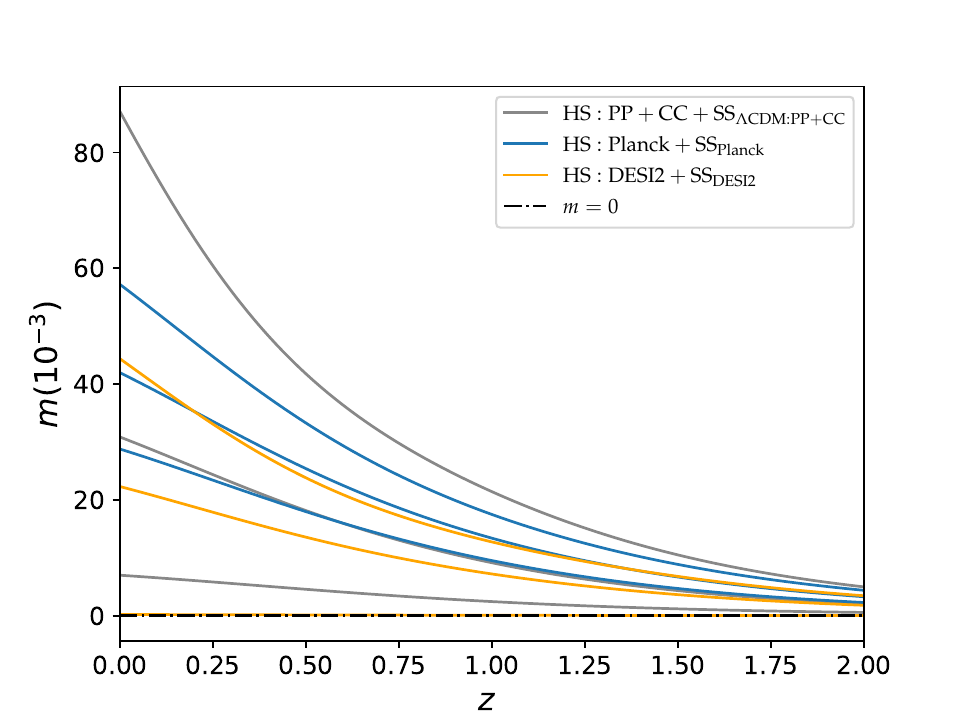}\\
	\includegraphics[width=4.1cm]{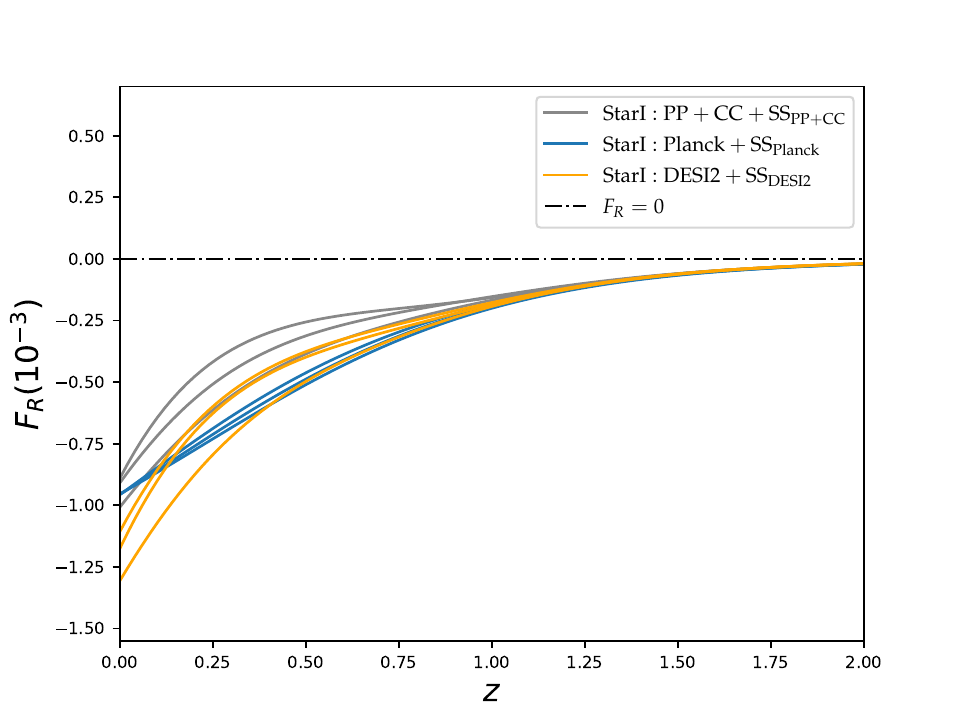}
	\includegraphics[width=4.1cm]{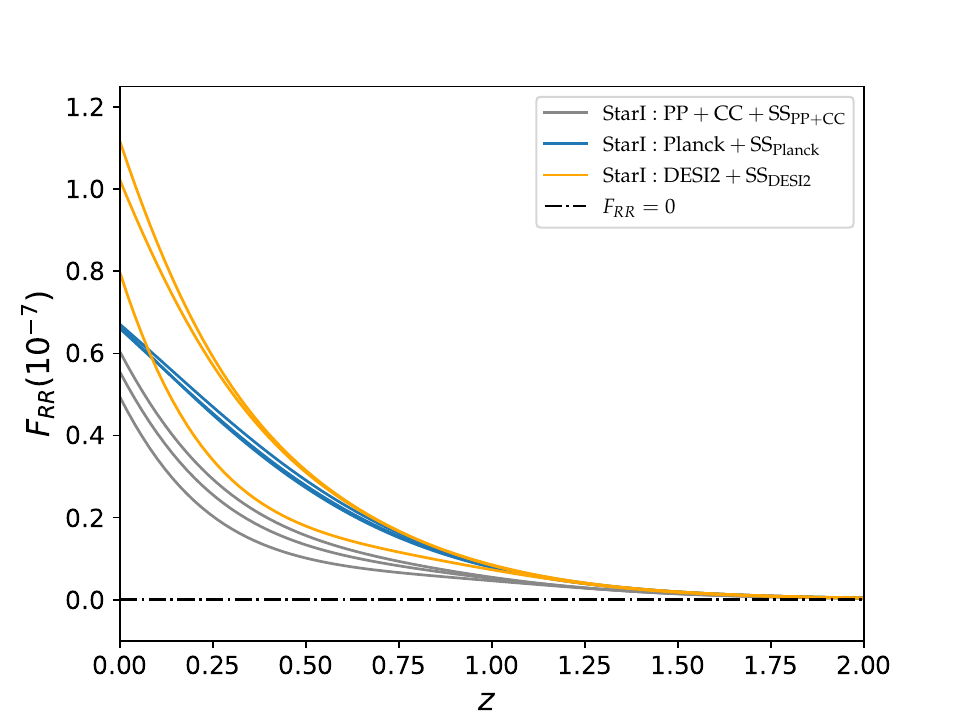}
	\includegraphics[width=4.1cm]{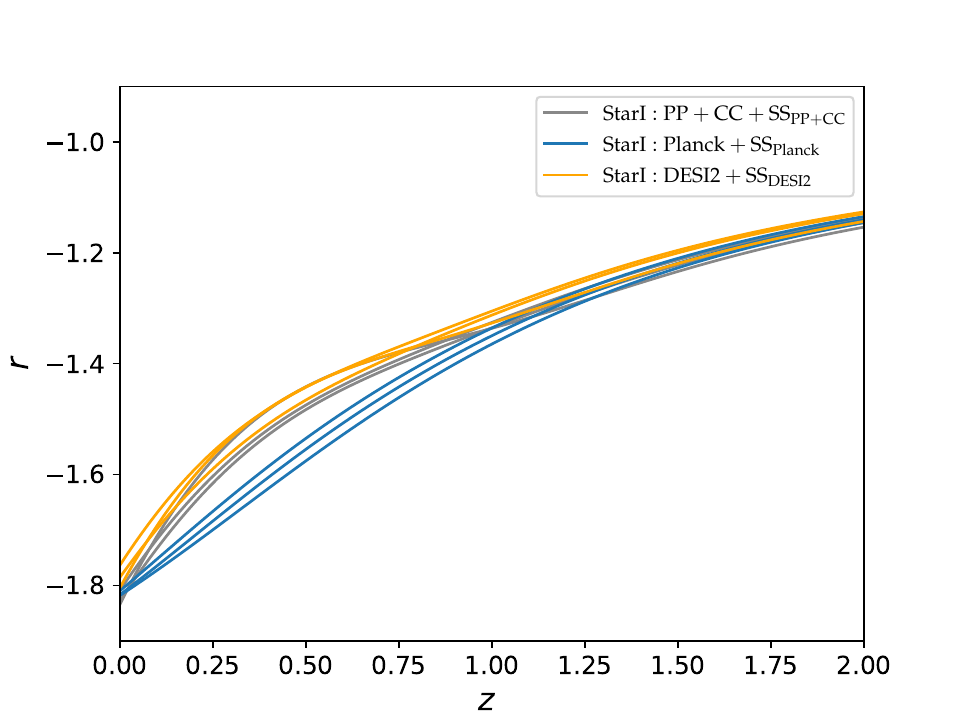}
\includegraphics[width=4.1cm]{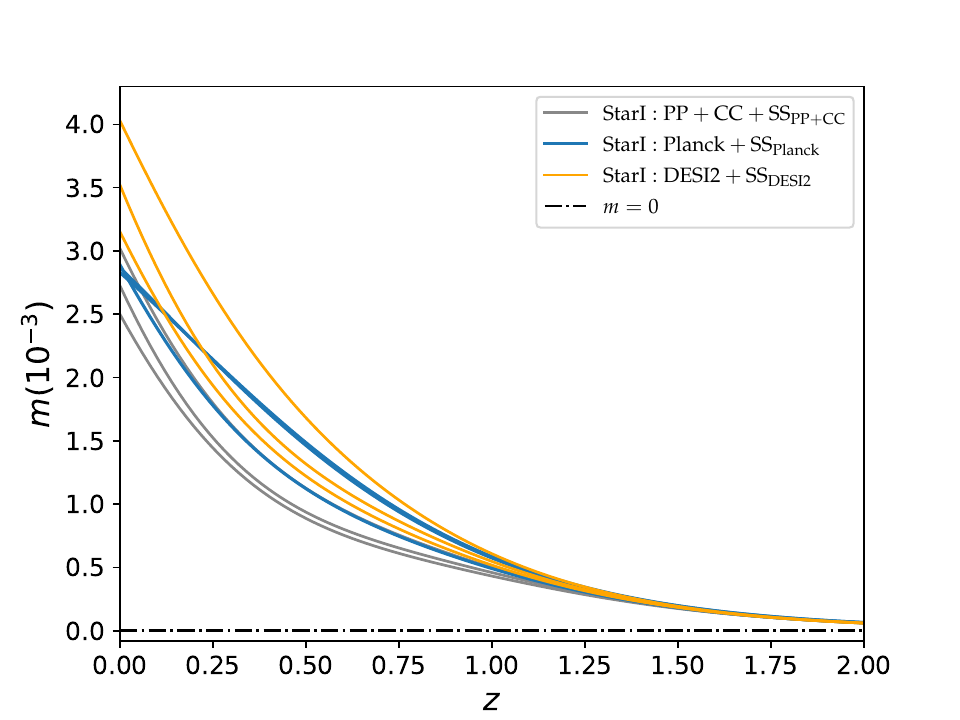}\\
	\includegraphics[width=4.1cm]{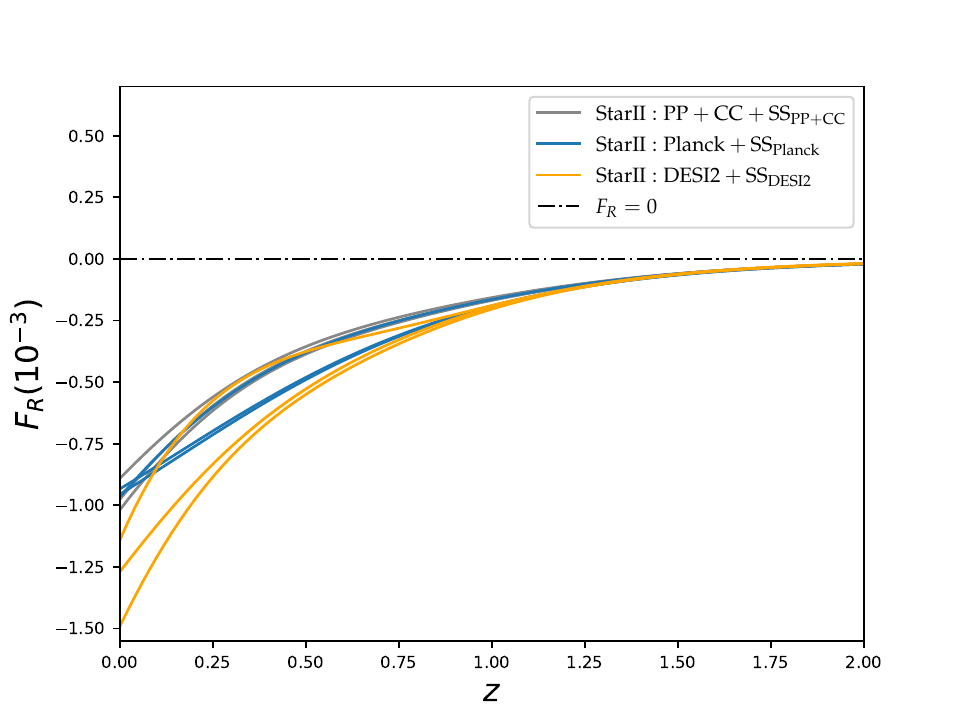}
	\includegraphics[width=4.1cm]{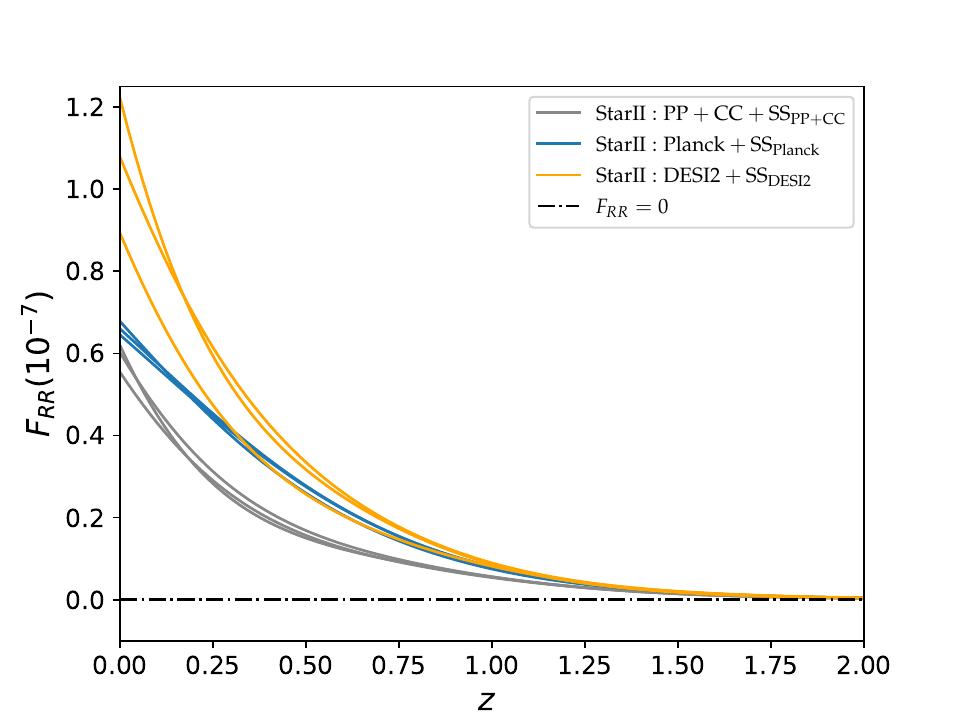}
	\includegraphics[width=4.1cm]{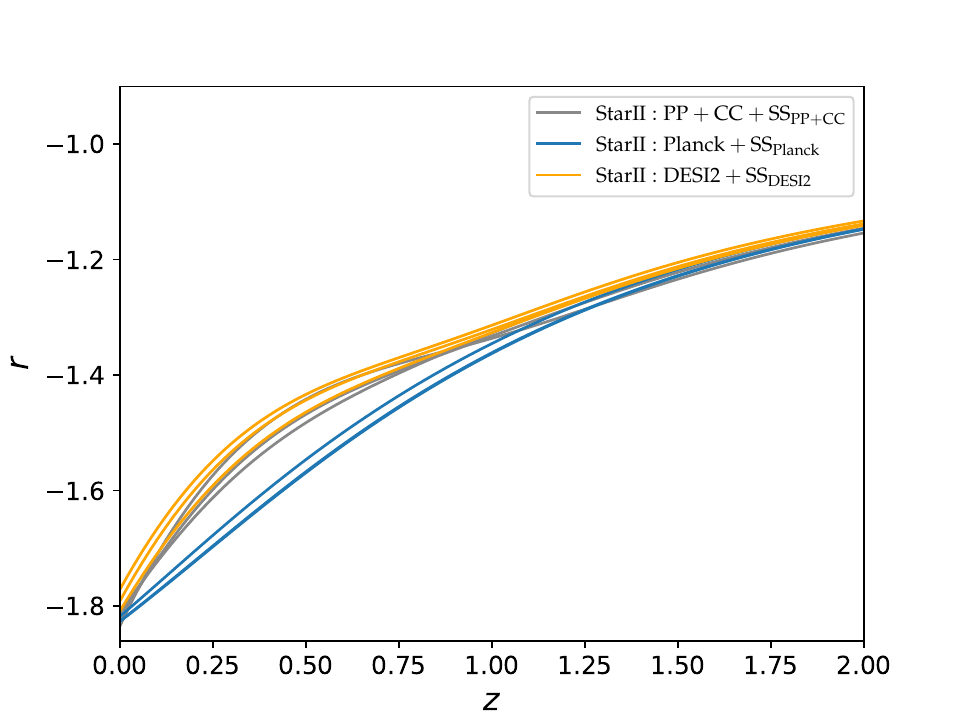}
\includegraphics[width=4.1cm]{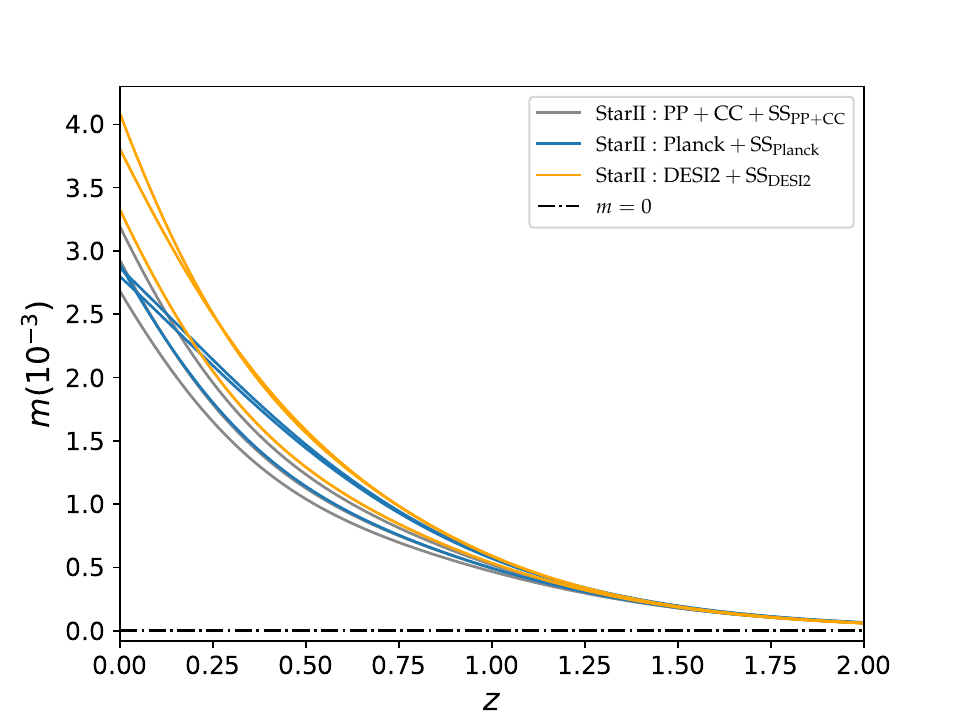}\\
	\caption{The evolutions of  $F_{R}$, $F_{RR}$, $r$, and  $m$   for the Hu-Sawicki (upper),  Starobinsky\Romannum{1} (middle) and Starobinsky\Romannum{2} (bottom)  models, which are based on the the $1\sigma$ regimes of PP+CC+SS$_{\rm \Lambda CDM:PP+CC}$/PP+CC+SS$_{\rm PP+CC}$,  DESI2+SS$_{\rm DESI2}$ and    Planck+SS$_{\rm    Planck}$ constraining results.}
	\label{frr}        
\end{figure*}
\section{Discussion}\label{Discussion}

The cosmological constraints are summarized in Tables~\ref{tfr} and \ref{taic}.
Generally, both the Hu-Sawicki and Starobinsky $f(R)$ models are consistent with
the data and can account for the late-time accelerated expansion.
For each dataset, the best-fit $\chi^2_{\min}$ is comparable to the number of
data points. In particular, the standard siren  dataset yields
$\chi^2_{\min}\simeq 1000$, indicating good internal consistency of the simulations.
As expected, the joint EM+SS analyses provide the tightest constraints for all
models, with the EM data contributing the dominant statistical weight.The impact of SS data on the Hubble tension is intrinsically dependent on the
adopted simulations and fiducial cosmology, since the $H_0$ values inferred
from SS analyses largely depend on the assumed fiducial model. Consequently, SS
datasets constructed with similar fiducial assumptions tend to yield comparable
levels of Hubble tension.

For the reference $\Lambda$CDM model, among the EM datasets, the Planck provides the
tightest constraints on $\Omega_{m0}$ and $H_0$.
The $1\sigma$ confidence regions from PP+CC, DESI2, and Planck are clearly
separated in the $\Omega_{m0}-H_0$ plane, in which the PP+CC constraint favors a
substantially larger $H_0$ that is lying outside the $2\sigma$ regions inferred from
Planck and DESI2.
Within $\Lambda$CDM, SS datasets based on Planck or DESI2
fiducials may mildly reduce the inferred tension, whereas
$\mathrm{SS}_{\mathrm{PP+CC}}$ favors a higher $H_0$ because it assumes a larger
fiducial Hubble constant. As a result, PP+CC+$\mathrm{SS}_{\mathrm{PP+CC}}$
yields a tension of $\sim 0.34\sigma$, compared with $3.82\sigma$ and
$6.06\sigma$ for the Planck and DESI2 related combinations, respectively,
within the adopted simulations.

As shown in Fig.~\ref{fr2tri}, for all considered $f(R)$ models Planck yields
the overall tightest constraints, while PP+CC-related combinations prefer
larger values of $H_0$ as $\Lambda$CDM model.
Figs.~\ref{fr1tri} and \ref{fr2tri} further show that our SS simulations are
in good agreement with the PantheonPlus sample.
For the Hu-Sawicki model, the SS luminosity-distance evolution follows a
similar ordering to that in $\Lambda$CDM,
\begin{eqnarray}
	\nonumber
	D_L^{\mathrm{SS}_{\mathrm{PP+CC}}}
	< D_L^{\mathrm{SS}_{\mathrm{DESI2}}}
	\lesssim D_L^{\mathrm{SS}_{\mathrm{Planck}}}.
\end{eqnarray}
In contrast, for the Starobinsky model (left panels of Fig.~\ref{fr2tri}) the
ordering becomes
\begin{eqnarray}
	\nonumber
	D_L^{\mathrm{SS}_{\mathrm{PP+CC}}}
	< D_L^{\mathrm{SS}_{\mathrm{Planck}}}
	\lesssim D_L^{\mathrm{SS}_{\mathrm{DESI2}}}.
\end{eqnarray}
In most cases, the total constraining power remains dominated by EM
observations.
The constraints on the Hu-Sawicki parameter $b$ are comparable to those in the
Starobinsky model; however, for the derived quantity $w_{\mathrm{eff}0}$ which
is mainly controlled by $b$, the Hu-Sawicki model allows a substantially
broader parameter range than the Starobinsky case.

Using the $1\sigma$ constraints from
the EM+SS joint data,
we show the redshift evolution of $w_{\mathrm{eff}}$ in Fig.~\ref{frw}, and of
$F_R$, $F_{RR}$, $r$ and $m$ in Fig.~\ref{frr}.
For all considered $f(R)$ models,the
effective equation of state $w_{\mathrm{eff}}$ exhibits mild oscillatory
behavior; $F_R$ and $r$ monotonically increase with
redshift $z$, whereas $F_{RR}$ and the deviation parameter $m$ decrease with $z$. All models smoothly approach the $\Lambda$CDM limit around
$z\simeq 2$, where $F_R\to 0$, $F_{RR}\to 0$, and $m\to 0$.
According to Eq.~(\ref{betaT}), the increase of $F_R$ implies $\beta_T\leq 1$,
which leads to a smaller SS luminosity distance $D_L^{\mathrm{SS}}$ compared to
its electromagnetic counterpart $D_L^{\mathrm{EM}}$.
And the Hu-Sawicki and Starobinsky models evolve
in a similar region of the $(r,m)$ phase space.


\subsection{Discussion:Hu-Sawicki model}

For the Hu-Sawicki model, all marginalized posterior probability density
functions are approximately Gaussian as shown in Fig.~\ref{fr1tri}. As reported in
Table~\ref{tfr}, the best-fit value $\chi^2_{\rm PP+CC}$ is slightly smaller
than that of the reference $\Lambda$CDM model. Taking $\Lambda$CDM as the
baseline, we obtain $|\Delta{\rm AIC}|<5$ in all cases, implying that the AIC
does not provide decisive model-selection evidence. By contrast, for the EM
datasets we find $\Delta{\rm BIC}=4.7/3.0/3.4$ for PP+CC, Planck, and DESI2,
respectively, which constitutes positive evidence against $\Lambda$CDM.
Moreover, since the PP+CC-based SS simulations adopt $\Lambda$CDM fiducial
parameters (see Sec.~\ref{method}), they also allow a meaningful assessment of
model-selection evidence within the simulated setup. Because the corresponding
SS constraints prefer values of $\Omega_{m0}$ that differ from those favored by
Planck and DESI2, the SS$_{\Lambda{\rm CDM:PP+CC}}$ related combinations yield
even larger BIC differences, $\Delta{\rm BIC}=7.4/8.3$. In the context of the
adopted simulations, this can be interpreted as strong evidence against
$\Lambda$CDM.

The constraints inferred from the SS simulations exhibit nearly parallel
confidence contours. Among them, SS$_{\rm DESI2}$ produces the widest
constraints. For instance, in the $\Omega_{m0}-H_0$ plane the relative
constraining power satisfies
\begin{eqnarray}
	\nonumber
	\mathrm{SS}_{\Lambda{\rm CDM:PP+CC}}
	< \mathrm{SS}_{\rm Planck}
	< \mathrm{SS}_{\rm DESI2}.
\end{eqnarray}
 In particular, PP+CC-related combinations yield
systematically smaller Hubble tensions than the Planck and DESI2 related
combinations which prefer lower values of $H_0$. The inclusion of SS data in
the Hu-Sawicki model can partially alleviate the inferred Hubble tension,
which is attributed to the modified GW friction term. Quantitatively, the
SS$_{\Lambda{\rm CDM:PP+CC}}$ combination gives a tension of $\sim 0.01\sigma$,
while the Planck- and DESI2-based SS simulations yield $\sim 0.41\sigma$ and
$\sim 2.44\sigma$ respectively.

Interestingly, as shown in Fig.~\ref{fr1tri}, the $w_{\rm eff0}-b$ contours
from the PP+CC and DESI2 related datasets exhibit a characteristic
``butterfly"-like morphology, where $w_{\rm eff0}$ and $b$ are positively correlated for
$b<0$ but are  negatively for $b>0$. Consistently,
Fig.~\ref{frw} shows that the effective equation-of-state parameter
$w_{\rm eff}(z)$ displays  oscillations with redshift. Also,
$w_{\rm eff}$ can cross the phantom divide and evolve from $w_{\rm eff}<-1$ to
$w_{\rm eff}>-1$ at the redshift range $0<z<2$, which may be associated with ghost-like instabilities in the
effective description.

For EM-only constraints, only PP+CC excludes $b=0$ (equivalently
$w_{\rm eff0}=-1$ and $m_0=0$) at the $1\sigma$ level, and excludes $F_{R0}=0$
(equivalently $F_{RR0}=0$ and $r_0=-2$) at the $2\sigma$ level. For the SS
simulations, only SS$_{\rm DESI2}$ yields a departure from $\Lambda$CDM
comparable to that obtained from PP+CC. The Planck constraints give a very
narrow range of $\Omega_{m0}$, and the $\Omega_{m0}-b$ and $H_0-b$ contours
show no significant correlation. In addition, for Planck-related combinations
$w_{\rm eff0}$ is relatively insensitive to $b$, leading to a broader allowed
range than in the other two cases. After combining Planck with
SS$_{\rm Planck}$, we find that $w_{\rm eff0}=-1$ is excluded at $1\sigma$, while
$b=0$ (together with $F_{R0}=0$, $F_{RR0}=0$, $r_0=-2$, and $m_0=0$) is excluded
at $2\sigma$. Notably, except for PP+CC, SS$_{\rm DESI2}$, and
Planck+SS$_{\rm Planck}$, the remaining constraints tend to prefer negative
values of $F_{RR0}$, which would imply theoretical instabilities.

Taking into account the preference for negative $F_{RR0}$, the crossing of
$w_{\rm eff}=-1$, and the BIC results, we conclude that the Hu-Sawicki model
may be disfavored and potentially ruled out by future high-precision
standard siren data.
\subsection{Discussion:Starobinsky model}

Since the Starobinsky model is an even function of the deviation parameter $b$,
we impose the prior $b\neq 0$. This choice artificially splits the parameter
space into two symmetric branches, $b<0$ and $b>0$, which we denote as
Starobinsky\Romannum{1} and Starobinsky\Romannum{2}, respectively. 
As shown in Table~\ref{tfr}, the best-fit $\chi^2$ values of the Starobinsky
model are smaller than those of $\Lambda$CDM for all datasets. In terms of model
selection, only the PP+CC dataset yields $\Delta{\rm BIC}=3.1$, corresponding to
weak positive evidence against $\Lambda$CDM. For the remaining datasets, the
information criteria do not provide decisive discrimination, indicating that
$\Lambda$CDM and the Starobinsky model are statistically comparable.

The PP+CC-related combinations systematically favor higher values of $H_0$ than
the Planck and DESI2 related combinations, thereby leading to a reduced Hubble
tension, similar to the other considered models. Relative to the
$5.87\sigma$ tension obtained from Planck alone in the Starobinsky\Romannum{1}
and Starobinsky\Romannum{2} branches, adding $\mathrm{SS}_{\rm Planck}$ mildly
alleviates the tension to $4.70\sigma/4.73\sigma$. In contrast, for PP+CC and
DESI2, the inclusion of SS data tends to increase the inferred tension in the
Starobinsky model. For instance, in the Starobinsky\Romannum{1} model, the
$\mathrm{SS}_{\rm PP+CC}$ and $\mathrm{SS}_{\rm DESI2}$ datasets yield tensions
of approximately $0.55\sigma$ and $9.07\sigma$ respectively, whereas the
corresponding EM datasets PP+CC/DESI2 give $\sim 0.11\sigma$ and
$1.88\sigma$ tensions, respectively.

Owing to the imposed prior $b\neq 0$, the posterior distributions for the
Planck-related combinations do not always exhibit Gaussian shapes (see
Fig.~\ref{fr2tri}). In particular, the $\mathrm{SS}_{\rm DESI2}$ posterior in
Starobinsky\Romannum{1} shows a clear non-Gaussian feature as well. Nevertheless, for
EM-only constraints  the two branches remain approximately mirror
symmetric, again reflecting the evenness of the model. The
$\Omega_{m0}$-$H_0$ contours inferred from PP+CC, Planck, and DESI2 are clearly
separated, indicating substantial inconsistencies among the EM datasets.
Meanwhile, the SS simulations produce nearly parallel confidence contours,
although the sizes of the allowed regions vary. Among the SS cases,
 the overall
constraining power (as inferred from contour areas) follows
\begin{eqnarray}
	\nonumber
	\mathrm{SS}_{\rm DESI2} < \mathrm{SS}_{\rm PP+CC} \leq \mathrm{SS}_{\rm Planck}.
\end{eqnarray}
Compared with EM only constraints, adding SS data systematically shifts
$\Omega_{m0}$ to larger values and $H_0$ to smaller values. Despite the fact
that the SS simulations are generated using symmetric EM best-fit values, the
resulting SS constraints are not perfectly symmetric where  the
Starobinsky\Romannum{1} branch typically exhibits a larger deviation from
$\Lambda$CDM than Starobinsky\Romannum{2} for the same $|b|$. This suggests that standard siren
observables are sensitive to the detailed curvature dependence of the $f(R)$
function. Although the Starobinsky model is even in $b$, cosmological evolution probes
only the $R>0$ branch. And the background dynamics depend nonlinearly on $F_R$
and $F_{RR}$. For the same $|b|$, the $b<0$ branch typically approaches the
late-time de~Sitter attractor more slowly, which can lead to a larger
accumulated deviation in the luminosity distance measured by standard sirens.

The $\Omega_{m0}-b$ correlation further illustrates the potential of the
Starobinsky model to break parameter degeneracies in a data dependent manner.
Specifically, Planck and DESI2 related constraints show a negative correlation,
whereas PP+CC favors a positive correlation; the SS constraints also tend to
prefer a negative correlation. Although PP+CC alone provides the loosest bounds,
its combination with SS data becomes significantly tighter which demonstrates the
degeneracy breaking power of standard sirens within this model.

The parameter correlations in the $\Omega_{m0}-b$, $H_0-b$, and
$w_{\rm eff0}-b$ planes shift signs between Starobinsky\Romannum{1} and
Starobinsky\Romannum{2} because of the mirror symmetry. In particular, $w_{\rm eff0}$ and $b$ are negatively
correlated for $b<0$, but positively correlated for $b>0$, producing a
``butterfly''-like structure analogous to that found in the Hu-Sawicki case.
At the same time, $w_{\rm eff}(z)$ evolves from the quintessence regime to the
phantom regime which is a stable behavior. Notably, the DESI2-related combinations can exclude the
$\Lambda$CDM limit $b=0$ (equivalently $w_{\rm eff0}=-1$) at the $2\sigma$ level,
providing evidence for dynamical dark energy within the Starobinsky framework.
For the Starobinsky model, we constrain $F_{R0}$ at the level of $\mathcal{O}(10^{-3})$,
tighter than that in the Hu-Sawicki model (typically $\mathcal{O}(10^{-2})$), and we
obtain $F_{RR0}\sim \mathcal{O}(10^{-7})$, which is positive and also smaller
than the Hu-Sawicki case (typically $\mathcal{O}(10^{-6})$). Moreover, all data
combinations exclude $F_{R0}=0$, $F_{RR0}=0$, $r_0=-2$, and $m_0=0$ at the
$2\sigma$ level for the Starobinsky model.
We find that the redshift evolution of $w_{\rm eff}$ in the combined DESI2 data, together with the
trajectories of $(F_R,F_{RR},r,m)$, can clearly distinguish the Starobinsky
scenario from $\Lambda$CDM model.
\section{Conclusion}\label{con}

In this work, we first constrained the Hu-Sawicki and Starobinsky $f(R)$ models with electromagnetic  datasets (PP+CC, Planck, and DESI2) and then simulated standard siren  catalogs based on the constrained results. Both $f(R)$ scenarios provide global fits comparable to $\Lambda$CDM, and the combined EM+SS analyses yield the tightest constraints.

The simulated SS data offer complementary sensitivity via the modified GW friction term, thereby enhancing the ability to discriminate $f(R)$ gravity from $\Lambda$CDM. However, in our simulation-based setup, the inferred Hubble tension is largely dominated by the assumed SS fiducial cosmology. Consequently, while standard sirens can help distinguish modified gravity from $\Lambda$CDM, they do not provide a viable solution to the Hubble tension in this context. 

For the Hu-Sawicki model, SS data can mildly reduce the tension, but several data combinations prefer $F_{RR0}<0$, suggesting potential theoretical instabilities. We conclude that the Hu-Sawicki model may be disfavored and potentially ruled out by future high-precision standard siren data. For the Starobinsky model, EM-only constraints are nearly symmetric between the $b<0$ and $b>0$ branches, whereas SS constraints show mild asymmetries, indicating sensitivity to the curvature dependence of $f(R)$.  The redshift evolution of $w_{\rm eff}$ in the combined DESI2 dataset, together with the trajectories of $(F_R, F_{RR}, r, m)$, clearly distinguishes the Starobinsky scenario from the $\Lambda$CDM model. Given the current inconsistencies in the data, future standard siren observations will be essential for a definitive assessment of $f(R)$ gravity as an alternative to the standard cosmological paradigm.

\section{Acknowledgements}
YZ is supported by National Natural Science Foundation of China under Grant No.12275037 and 12275106. DZH is supported by the Talent Introduction Program of Chongqing University of Posts and Telecommunications (Grant No.E012A2021209), the Youth Science and technology research project of Chongqing Education Committee (Grant No.KJQN202300609).

%

\bibliography{frtex}
\bibliographystyle{utphys}

\end{document}